\documentclass[useAMS,usenatbib]{mn2e}
\usepackage{amsmath}
\usepackage{amssymb}
\usepackage{graphicx}
\usepackage{pdflscape}
\usepackage{xcolor}

\title[NGTS Autovetting]{Automatic vetting of planet candidates from ground based surveys: Machine learning with NGTS}

\author[Armstrong et. al.]{
\parbox{\textwidth}{David. J. Armstrong,$^{1,2}$\thanks{d.j.armstrong@warwick.ac.uk}
Maximilian~N.~G{\"u}nther,$^{3}$
James McCormac,$^{1,2}$
Alexis~M.~S.~Smith,$^{4}$
Daniel~Bayliss,$^{1,2}$
Fran\c{c}ois Bouchy,$^{5}$
Matthew~R.~Burleigh,$^{6}$
Sarah~Casewell,$^{6}$
Philipp~Eigm\"uller,$^{4}$
Edward~Gillen,$^{3}$
Michael~R.~Goad,$^{6}$
Simon~T.~Hodgkin,$^{7}$
James~S.~Jenkins,$^{8,9}$
Tom Louden,$^{1,2}$
Lionel Metrailler,$^{5}$
Don~Pollacco,$^{1,2}$
Katja Poppenhaeger,$^{10}$
Didier~Queloz,$^{3}$
Liam~Raynard,$^{6}$
Heike~Rauer,$^{4,11}$
St\'{e}phane~Udry,$^{5}$
Simon.~R.~Walker,$^{1}$
Christopher~A.~Watson,$^{10}$
Richard~G.~West,$^{1,2}$
Peter~J.~Wheatley$^{1,2}$
}
\vspace{0.4cm}\\
\parbox{\textwidth}{
$^{1}$University of Warwick, Department of Physics, Gibbet Hill Road, Coventry, CV4 7AL, UK\\
$^{2}$Centre for Exoplanets and Habitability, University of Warwick, Gibbet Hill Road, Coventry CV4 7AL, UK\\
$^{3}$Astrophysics Group, Cavendish Laboratory, J.J. Thomson Avenue, Cambridge CB3 0HE, UK\\
$^{4}$Institute of Planetary Research, German Aerospace Center, Rutherfordstrasse 2, 12489 Berlin, Germany\\
$^{5}$Observatoire de Gen{\`e}ve, Universit{\'e} de Gen{\`e}ve, 51 Ch. des Maillettes, 1290 Sauverny, Switzerland\\
$^{6}$Department of Physics and Astronomy, Leicester Institute of Space and Earth Observation, University of Leicester, LE1 7RH, UK\\
$^{7}$Institute of Astronomy, University of Cambridge, Madingley Rise, Cambridge CB3 0HA, UK\\
$^{8}$Departamento de Astronomia, Universidad de Chile, Casilla 36-D, Santiago, Chile\\
$^{9}$ Centro de Astrof\'isica y Tecnolog\'ias Afines (CATA), Casilla 36-D, Santiago, Chile.\\
$^{10}$Astrophysics Research Centre, School of Mathematics and Physics, Queen's University Belfast, BT7 1NN Belfast, UK\\
$^{11}$Center for Astronomy and Astrophysics, TU Berlin, Hardenbergstr. 36, D-10623 Berlin, Germany
}}

\newcommand{\mytilde}{\raise.17ex\hbox{$\scriptstyle\mathtt{\sim}$}}

\begin{document}
\date{Accepted . Received}

\pagerange{\pageref{firstpage}--\pageref{lastpage}} \pubyear{2002}

\maketitle

\begin{abstract}
State of the art exoplanet transit surveys are producing ever increasing quantities of data. To make the best use of this resource, in detecting interesting planetary systems or in determining accurate planetary population statistics, requires new automated methods. Here we describe a machine learning algorithm that forms an integral part of the pipeline for the NGTS transit survey, demonstrating the efficacy of machine learning in selecting planetary candidates from multi-night ground based survey data. Our method uses a combination of random forests and self-organising-maps to rank planetary candidates, achieving an AUC score of 97.6\% in ranking 12368 injected planets against 27496 false positives in the NGTS data. We build on past examples by using injected transit signals to form a training set, a necessary development for applying similar methods to upcoming surveys. We also make the \texttt{autovet} code used to implement the algorithm publicly accessible. \texttt{autovet} is designed to perform machine learned vetting of planetary candidates, and can utilise a variety of methods. The apparent robustness of machine learning techniques, whether on space-based or the qualitatively different ground-based data, highlights their importance to future surveys such as TESS and PLATO and the need to better understand their advantages and pitfalls in an exoplanetary context.
\end{abstract}


\section{Introduction}
The detection of exoplanets through photometric observation of transits has driven the field in recent years. Progress has been made from hard fought discoveries of single giant planets to large scale surveys capable of finding thousands of Earth, Neptune and Jupiter sized planets \citep{Charbonneau:2000fh,Bouchy:2005wb,CollierCameron:2007ew,Bakos:2007tt,Borucki:2010dn,Morton:2016ka,Thompson:2017wu}. Future progress will focus on a combination of single rare or interesting systems alongside large scale population studies highlighting trends in the distribution and occurrence of planets.

A typical search for transiting planets might follow the process of 1) detrending the data \citep[e.g.][]{Stumpe:2012bj,Smith:2012ji,Tamuz:2005hi}, 2) running a search algorithm for transiting planetary signals \citep{2006MNRAS.373..799C,Kovacs:2002ho,Mislis:2015cj,Pearson:2018kf}, 3) vetting the results to produce a candidate list \citep[e.g. \textit{Kepler}'s Robovetter,][]{Coughlin:2016cm}, and 4) following up or validating these candidates to find new planets \citep{Diaz:2014kd,Morton:2012bv,Santerne:2015bb,2015ApJ...800...99T}. Step 3, vetting, often involves significant human input, whether by `eyeballing' each significant signal or by setting thresholds on a series of semi-automated tests. This is especially true when considering ground based surveys \citep{Bakos:2004gx,Bakos:2013fc,McCullough:2005do,Alonso:2004ii,Pollacco:2006gb,Pepper:2007ja,Wheatley:2017dm}, which must typically deal with more complex window functions and atmospheric noise sources than a space-based survey. Methods have been published to automate vetting in the past, with \citet{McCauliff:2015fb} applying random forests to classify \textit{Kepler} candidates, \citet{Thompson:2015ic} and \citet{Armstrong:2017cp} clustering \textit{Kepler} and K2 candidate lightcurves with similar shapes and using the results to classify candidates, and \citet{Shallue:2017vy} applying neural nets to the classification of Kepler TCEs into planet candidates or false positives. These methods aim to build towards a situation where the process of planet detection can be fully automated, with the long term goal of combining steps 3 and 4; although planet validation on a large scale has been performed \citep{Crossfield:2016ip,Morton:2016ka}, this relies on significant human input prior to and during the process, and there is the potential for problems if care is not taken on individual candidates \citep{Cabrera:2017fo,Shporer:2017hv}. 

Automation is both desirable and necessary. Future surveys will produce quantities of lightcurves and candidates beyond the scope of most methods. The soon to be launched TESS satellite \citep{Ricker:2014fy} is expected to observe tens of thousands of planets and hundreds of thousands of false positives, among $10^8$ targets \citep{Sullivan:2015ey}. Far more apparently significant signals will need rejecting, as a result of instrumental or stellar noise. Robustly automating the process will be necessary for dealing with this data quantity, and crucial for considering statistical properties of the planet population. Testing and removing human bias in the selection process is non-trivial, and repeatable automated methods allow for sensitivity testing and quantifiable debiasing. 

The only machine learning work in the exoplanet field applied to ground-based data to date is \citet{Dittmann:2017df}, who used a neural net to identify `trigger' events from single images of MEarth data. In this paper we present a technique to rank planetary candidates from the NGTS survey \citep{Wheatley:2017dm}, using ground-based data taken over several months for 46470 target objects, subject to the usual weather and visibility constraints. As such, the window function and noise properties of the data are among the most complex to have machine learning applied in an exoplanetary context. The method builds on that trialled on \textit{Kepler} in \citet{McCauliff:2015fb}, with different features targeting ground-based data, and the addition of a self-organising-map to characterise signal shape efficiently. We aim to improve the survey pipeline through automatic ranking and selection of candidates, while demonstrating the method's applicability to an increased quantity of data with complex noise sources, in preparation for future surveys. Here we focus on the vetting procedure (the step from significant signals to planetary candidates) but the method has the potential to be expanded to include validation (planetary candidates to validated planets) in future.

We draw established techniques from the machine learning field, utilising random forests \citep{Breiman:fb} and self-organising-maps \citep{Kohonen:1982dy}. Each has been demonstrated in an astrophysical context, in the realm of classifying stellar variability \citep{Eyer:2005ce,Mahabal:2008if,Blomme:2010bq,Debosscher:2011kz,Brink:2013hv,Nun:2014kv,Richards:2012ea,Brett:2004cr,Masci:2014bk,Farrell:2015je}, quasars \citep{Carrasco:2015kh}, redshifts \citep{CarrascoKind:2014gb}, solar flares \citep{Liu:2017dv} and asteroids \citep{Huang:2017eb}, to name but a few.


This paper is structured as follows. In Section \ref{sectNGTS} we describe the NGTS survey and the context of this method in the pipeline. In Section \ref{secttrainingset} we describe our training sets and testing strategy for the models. In Section \ref{sectMeth} we describe our methodology and hyperparameter choices, followed by Results in Section \ref{sectResults}, tests to guard against bias and overfitting in Section \ref{sectVal}, and we conclude in Section \ref{sectConc}.

\section{The NGTS survey}
\label{sectNGTS}
\subsection{Overview}
NGTS is a facility dedicated to detecting super-Earth and greater sized planets, through providing photometry at extremely high precision from the ground. NGTS obtains photometry at 0.1\% precision in 1 hour on stars brighter than 13th magnitude through a combination of focused design and extremely stable autoguiding \citep{McCormac:2013hc,McCormac:2017ju}. With a red-sensitive filter, the mission is optimised for the detection of small planets around K and early M stars. In total, NGTS is predicted to find $\sim$300 new exoplanets and $\sim$5600 eclipsing binaries \citep{Guenther:2017a}. The NGTS facility consists of 12 20cm f/2.8 telescopes sited at Cerro Paranal, Chile, using back-illuminated deep-depletion CCD cameras. Data is taken using one of the telescopes for each field, with each telescope typically observing two fields per night. Fields are not typically reobserved after the end of an observing season. At the point of writing, 10s cadence lightcurves for 46470 targets for on average 500 hours each spread over 250 nights were available for development and testing. See \citet{Wheatley:2017dm} for a full description of the facility, data collection strategy and photometry reduction.

\subsection{Candidate Detection}
\label{sectCDetection}
Potential planetary transit signals are searched for using an implementation of the BLS algorithm, named \texttt{orion}, which has been used for the majority of transit detections in the WASP project. The most significant five periodogram peaks are extracted for each target, with each peak assigned a rank from 1 to 5 ordered by the BLS signal strength. \texttt{orion} removes periods within 5\% of 1d or 2d before selecting the top 5 peaks. Peak periods which occur on a large number of objects in the same field are then removed (see Section \ref{secttrainingset} for detail), including any remaining near further aliases of 1 day. This leads to up to 5 periodogram peaks being used as separate detected candidates for each observed target, with less if some were removed. The large majority of these peaks represent instrumental signatures, harmonics, or other non-planetary signals. 

\subsection{Context}
Typically, the strongest signal in each lightcurve would be checked by a team of researchers looking for plausible planetary transits, and the best signals passed on for further photometric or spectroscopic follow-up. This human eyeballing is time intensive and can lead to complex biases in the signals selected, particularly for shallow marginal candidates, while also ignoring secondary BLS peaks on a target. The aim of this paper is to take steps towards automating this process, using machine learning algorithms to initially rank the candidate signals seen by researchers in order of their likelihood to be a planet, and with the eventual goal of removing human selection from the process entirely. Similar techniques have shown demonstrable success using space-based lightcurves; here we wish to extend the technique to the often more challenging ground-based data. The procedure implemented and tested here forms part of the data processing pipeline for NGTS, and is currently used to rank candidates before they are seen by researchers. After further testing and development ranking may extend to actual selection of targets for followup.

\section{Training and Test Sets}
\label{secttrainingset}
\subsection{Training}
`Supervised' machine learning techniques require a training set, a distribution of examples designed to teach the algorithm how certain classes of signal appear. In a mature survey, a set of already detected planets and false positives could be used. However, for practical use it is necessary to implement these algorithms at an earlier stage, where no such set exists for the survey in question. Planets detected using other facilities have different quantities and qualities of data available, and have survey dependent parameter distributions, complicating their use. As such we turn to synthetic, injected planets. 

Initially, we must generate model parameters for the star and injected planet. We randomly select actually observed NGTS targets, limiting the stellar magnitude from 9-14. Stellar effective temperature is then estimated using the 2MASS J-H colour for these targets, following \citet{CollierCameron:2007cy}, through

\begin{equation}
T_\textrm{eff} = -4369.5\left(J-H\right)+7188.2
\end{equation}

with stellar radius $R_*$ estimated from $T_\textrm{eff}$ through

\begin{multline}
\frac{R_*}{R_\odot} = -3.925\times10^{-14}(T_\textrm{eff})^4+8.3909\times10^{-10}(T_\textrm{eff})^3\\-6.555\times10^{-6}(T_\textrm{eff})^2+0.02245(T_\textrm{eff})-27.9788
\end{multline}

and stellar mass $M_*$ estimated as

\begin{equation}
\frac{M_*}{M_\odot} = \left(\frac{R_*}{R_\odot}\right)^{-0.8}
\end{equation}

These relations are valid for single main sequence stars with $4000\textrm{K}<T_\textrm{eff}<7000\textrm{K}$ with a scatter around the relation of order 100K \citep{CollierCameron:2007cy}. Transit parameters are drawn as uniform in log radius ratio between 0.1 and 2\%, uniform in orbital period between 0.35 and 20d, and uniform in impact parameter, allowing for grazing transits. We inject transits into 6 observing fields, and verify later that this does not bias results on the other fields. Model transits are injected into NGTS lightcurves prior to detrending. As such any effects of the detrending pipeline should be incorporated in the final lightcurve. Modified lightcurves are then searched for transits using the same BLS implementation as for the normal survey, and only those lightcurves where the injected signal was detected are put forward for use in our training set. As such the distribution of synthetic transits is affected by the NGTS sensitivity profile, and only transits which we could plausibly detect are used for training.

A training set is similarly needed for false positives, signals which may be flagged by the BLS algorithm but are not planetary in origin. In this instance such signals can include both instrumental artefacts (data gaps, remnant trends, etc) and astrophysical false positives (e.g. contaminating eclipsing binaries). The ideal training set for this group is the real detected candidates themselves. By using these candidates, we have the exact distribution of signals produced by the algorithm. Of course, within this set are some real planets, which represent the overall targets of the survey. Hence, we are proceeding under the assumption that the large majority of flagged signals are not planetary in origin and as such overwhelm the true planets present.

To limit the computational resources required, for both sets we apply a cut, removing candidate peaks which a) match another peak's period on the same object to within 0.2\% (removing all but the strongest peak) or b) match a period which shows an abnormal prevalence within an object's observed field, with greater than 5 other candidates showing a period within 0.2\%. For context, fields had a median of 2354 candidates after removal of matching peaks on the same object, with fields ranging between 756 and 5708. The aim of the cuts is to remove both alias periods on integer days caused by the observing pattern, as well as systematic noise periods connected to a particular field. The combination of targets, detrending, data quality and window function for a field can lead to spurious BLS detections which are common to several stars. In the latest run of NGTS data on all observed fields, the period cuts reduce the number of candidates to study from 188588 to 34668 (noting that one target object can have up to 5 candidates). There is some risk in applying this cut, as real planets with orbital periods very close to integer days would typically be removed from all fields. An example is WASP-131b \citep{Hellier:2017jv}, which lies in an NGTS field but was not observed for long enough to obtain a full transit. However, such planets are extremely hard to detect from the ground regardless due to observing window aliases. We further remove any candidate where any of our calculated features returned a null result (typically poor quality lightcurves with very high noise or few datapoints).

We use two versions of the NGTS pipeline during the development and testing phases. These are known as `TEST18', which after the above cuts contains 20166 candidates and 11005 synthetic injections, and the more developed `CYCLE1706', which again after cuts contains 27496 candidates and 12368 injections.

\subsection{Testing}
Typically, the development of a machine learning algorithm incorporates initial training on the training set, development of model parameters and other meta-choices using cross-validation on this training set, and final testing on an independent, not before seen test set. We adapt this to our survey by performing initial training and development on one iteration of the NGTS pipeline (TEST18), using cross-validation to test and verify feature selection and model parameters. We then test the models on a later version of the pipeline which became available during development (CYCLE1706). Note that in the end, models are trained using CYCLE1706 data and injections, and tested using cross-validation - but no model parameter choices or developmental decisions are made at this point, and hence the effect will be the same as the standard method. We appreciate that this is an unorthodox method for ensuring testing validity, but closely matches how the models will be used in practice; pipeline versions are continuously updated, and at each stage the models will be re-trained and applied on the new pipeline data, without new model development. To reinforce our test, we also perform a more typical training-test set analysis where the models are trained on the bulk of CYCLE1706 data then tested using a single field of CYCLE1706 data not before seen by the classifier, which produced similar results although with more limited numbers. The scores from these tests are presented in Section \ref{sectResults}.

\section{Methodology}
\label{sectMeth}
\subsection{Choice of techniques}
A wide array of machine learning techniques have been explored in the literature, with popular methods including deep neural nets \citep{LeCun:2015dt,Shallue:2017vy} and a variety of ensemble classifiers including Random Forests \citep[][hereafter RFs]{Breiman:fb}. Here we opt to use a combination of RFs with an unsupervised method, the self-organising-map \citep[hereafter SOM]{Kohonen:1982dy}. This combination has been demonstrated successfully in the past for classifying variable stars \citep{Armstrong:2016br} and the separate methods have both been demonstrated in a transiting exoplanet context \citep{McCauliff:2015fb,Armstrong:2017cp}. RFs have the advantage of being robust to varying data gaps and durations (field-to-field for example), as they deal with features extracted from the lightcurve rather than the lightcurve itself. RFs are also naturally extendable to include additional information, such as GAIA derived stellar parameters in the future, and in principle can be applied directly to other surveys or datasets to search for planets, although experience suggests this is rarely simple for any method. Finally, RFs allow a degree of interpretability, in terms of showing which features have the most deterministic power; this is a useful property for finding areas of potential weakness and bias in our models, and understanding how potential candidates are viewed by the model.

\subsection{SOM}
A SOM is a form of unsupervised machine learning, not requiring any training set. A SOM will cluster groups of inputs based on their proximity to each other, defining that proximity as the Euclidean distance between input points. We give an overview of the technique here; for more detail we refer the reader to \citet{Armstrong:2016br} for the setup and training methodology and \citet{Armstrong:2017cp} for application to exoplanets and extraction of the features used. 

The SOM used here consists of a 20x20 grid of `pixels', with each pixel consisting of a template transit. The grid is known as the Kohonen layer. The pixel templates are initialised randomly; as the SOM is trained they are permuted to resemble different shapes in the inputs. To train the SOM we use binned phase-folded lightcurve shapes, using periods, epochs and transit durations output from the transit search. Phase-folded lightcurves were cut to a window covering three transit durations, centred on the candidate transit, then binned into 20 equal-width bins. As such each SOM pixel has 20 values associated with it. Bin values were the weighted mean of all datapoints in the bin, using inverse square flux errors as the weights. We then normalise each transit such that the bottom of the transit has flux value 0, and the baseline value 1, measured using the lowest and highest quarter of points respectively. As such depth and duration information is removed. We randomly downsampled the real candidates to match the synthetic injections such that 11005 examples were available for each class, to give balanced inputs. 

The training process then proceeds as follows. The input parameters are the initial learning rate, $\alpha_0$, which influences the rate at which pixels in the Kohonen layer are adjusted, and the initial learning radius, $\sigma_0$, which affects the size of groups that emerge. Initially each pixel is randomised so that each of its 20 elements lies between 0 and 1, as our binned transits have been scaled to this range. For each of a series of iterations, each input is compared to the Kohonen layer. The best matching pixel in the layer is found, via minimising the Euclidean distance between the pixel elements and the input. Each element in each pixel in the layer is then updated according to the expression

\begin{equation}
m_{xy,k,new} = \alpha e^\frac{-d_{xy}^2}{2\sigma^2}  \left(s_k - m_{xy,k,old}\right)
\end{equation}

\noindent where $m_{xy,k}$ is the value $m$ of the pixel at coordinates $x$,$y$ and element $k$ in the binned transit, $d_{xy}$ is the euclidean distance of that pixel from the best matching pixel in the layer, and $s_k$ is the kth element of the considered input transit. Note that distances are continued across the Kohonen layer boundaries, i.e. they are periodic. Once this has been performed for each phase curve, $\alpha$ and $\sigma$ are updated according to

\begin{equation}
\label{eqnsigmadecay}
\sigma = \sigma_0 e^{\left(\frac{-i*log(r)}{n_\textrm{iter}}\right)}
\end{equation}
\begin{equation}
\label{eqnalphadecay}
\alpha = \alpha_0 \left( 1 - \frac{i}{n_\textrm{iter}}   \right)
\end{equation}

\noindent where $i$ is the current iteration, and $r$ is the size of the largest dimension of the Kohonen layer. This is then repeated for $n_\textrm{iter}$ iterations. 

It is possible to use different functional forms for the evolution of $\alpha$ and $\sigma$; typically a linear or exponential decay is used. \citet{Brett:2004cr} found that the performance of the SOM was largely unimpeded by the choice of form or initial value, as long as the learning rate does not drop too quickly. We find satisfactory results for the expressions above and values of $\alpha_0=0.1$ and $\sigma_0=20$, using $n_\textrm{iter}=300$ iterations.

\begin{figure*}
\resizebox{\hsize}{!}{\includegraphics{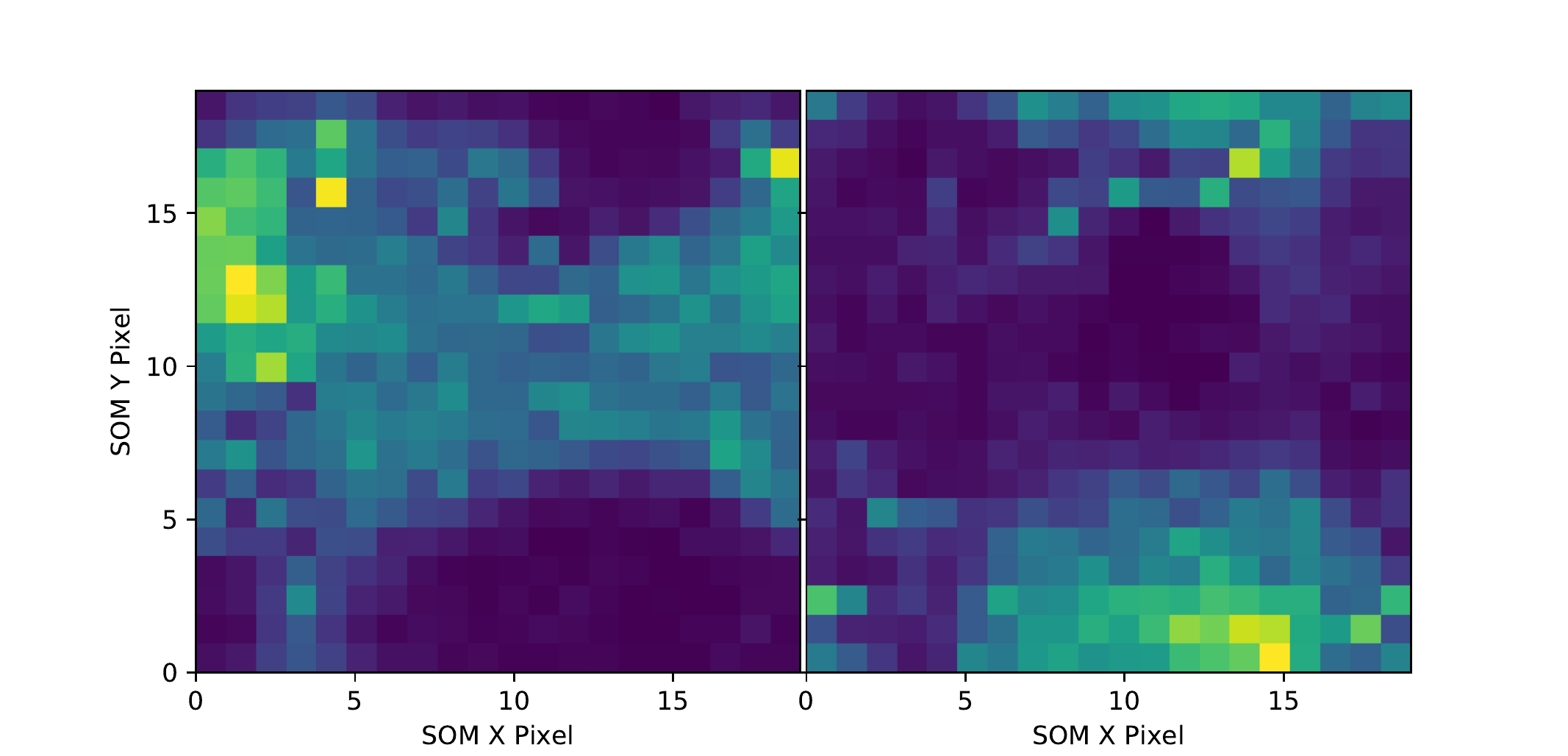}}
\caption{Histograms of positions on the trained 2 dimensional SOM for real candidates (left) and synthetic transit injections (right). Note the separation of the two groups.}
\label{figsomhist}
\end{figure*}

\begin{figure}
\resizebox{\hsize}{!}{\includegraphics{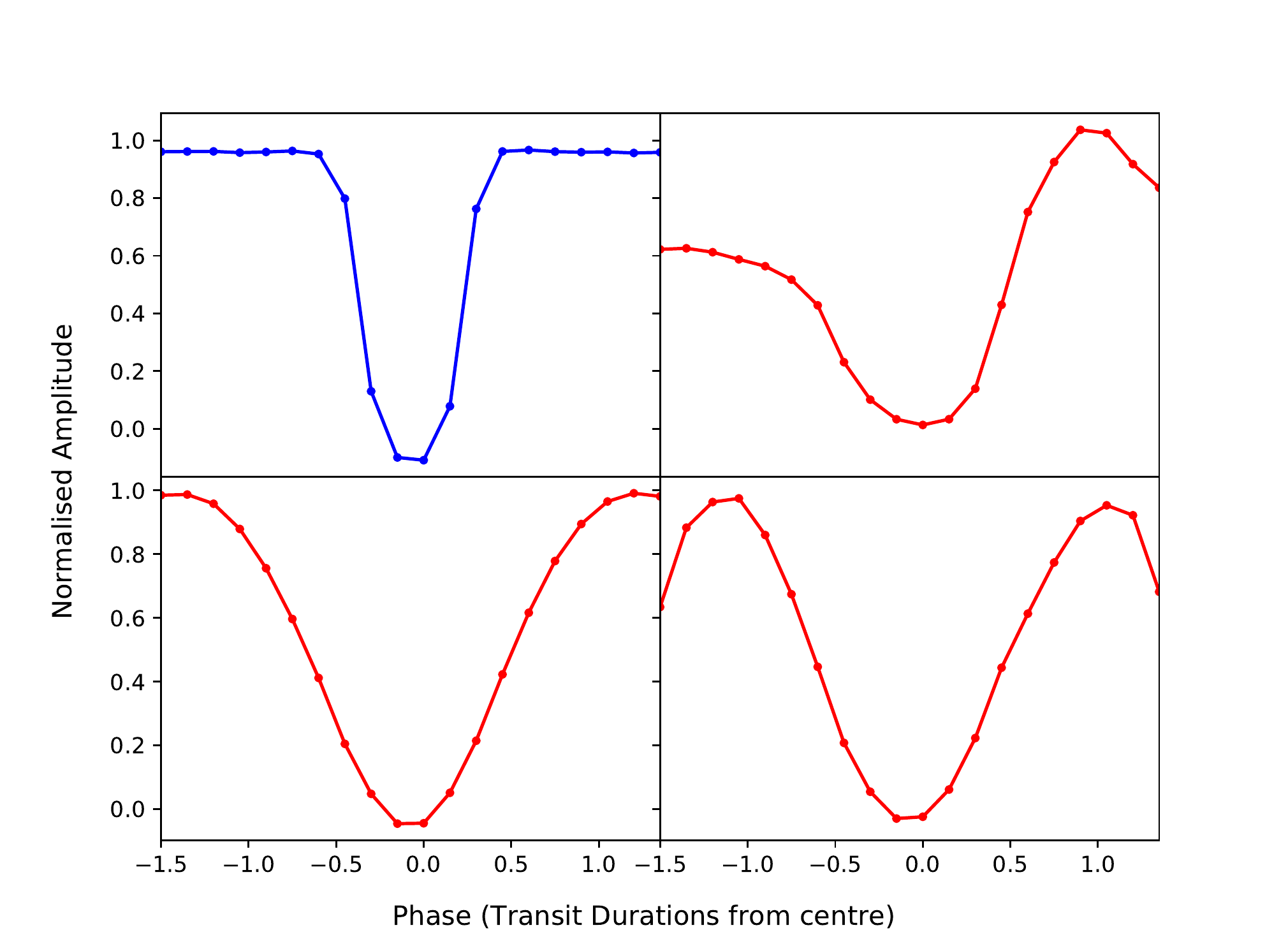}}
\caption{The SOM pixel templates for four key locations in Figure \ref{figsomhist}, one planetary (blue) and three non-planetary (red). Clockwise from top left the pixel indices are [15,0], [16,11], [4,16], [1,16] ([x,y] format).}
\label{figsomtemplates}
\end{figure}

\begin{figure}
\resizebox{\hsize}{!}{\includegraphics{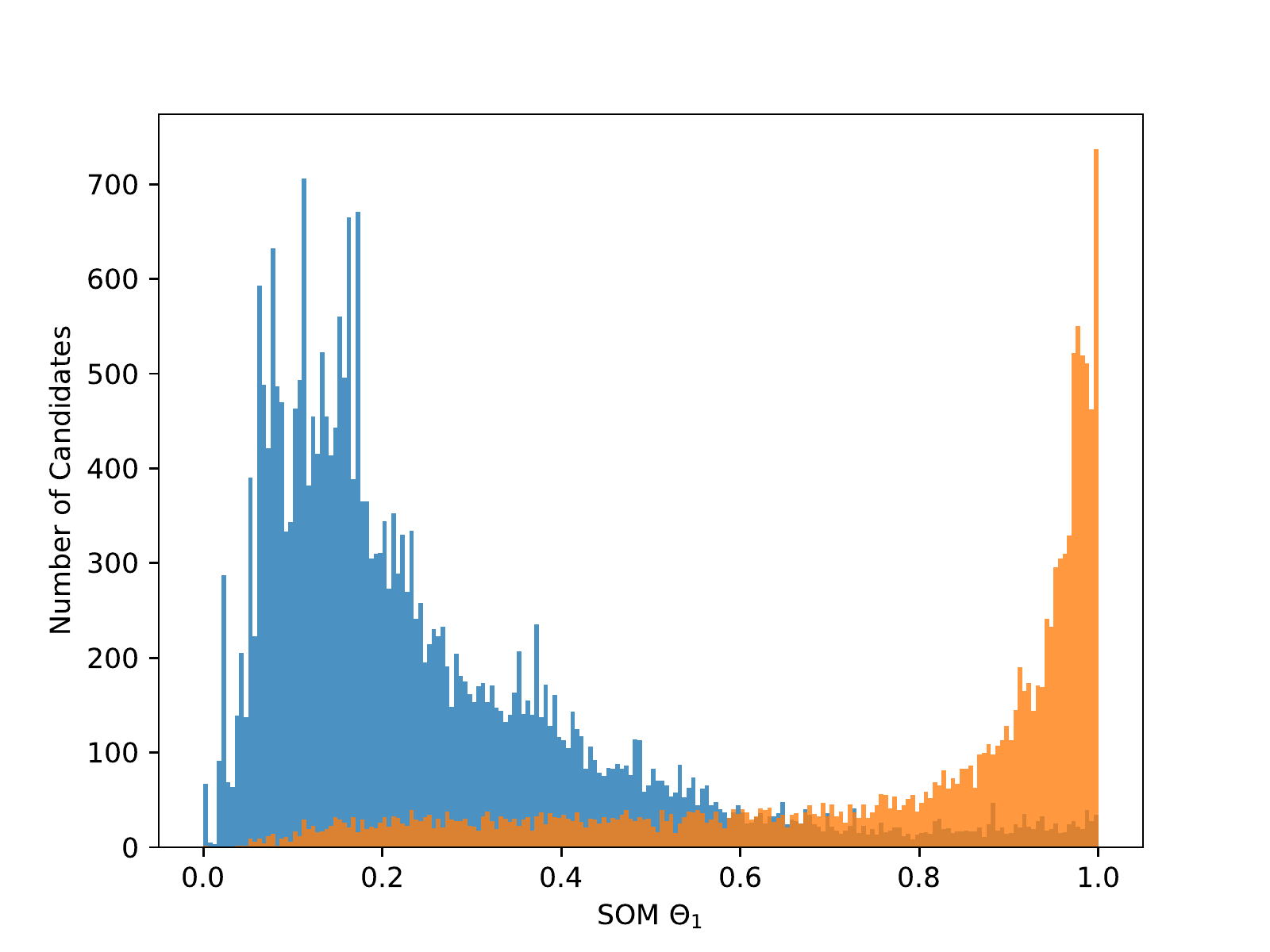}}
\caption{Distribution of $\theta_1$ for real candidates (blue) and synthetic transit injections (orange). The clustering evident in Figure \ref{figsomhist} leads to distinct distributions in $\theta_1$.}
\label{figtheta1dists}
\end{figure}

The resulting distribution of candidates in the trained SOM is shown in Figure \ref{figsomhist}, and example trained pixel templates in Figure \ref{figsomtemplates}. The degree of clustering shown demonstrates that the method is still powerful even on the very different candidates arising from a ground based survey, as compared to \emph{Kepler} where it was previously tested. In this case, rather than separating planetary candidates from false positives such as blended eclipsing binaries, the SOM is separating planetary candidates from a wide range of noise sources, astrophysical and instrumental, presented by the data. The location of a given candidate on the SOM is converted into the statistic $\theta_1$, which is described in \citet{Armstrong:2017cp}. The distribution of $\theta_1$ for each class is shown in Figure \ref{figtheta1dists}, and as expected from the observed clustering proves a powerful diagnostic in its own right.

The code used to implement the SOM can be found at \footnote{https://github.com/DJArmstrong/TransitSOM}, and is an extension of the SOM incorporated in the PyMVPA Python package \citep{Hanke:2009bm}.

\subsection{Random Forest}
Random Forests \citep{Breiman:fb} are a versatile and effective machine learning technique. A RF classifier uses an ensemble of Decision Trees to perform classifications, using a set of input features. In our case features are extracted from the lightcurves and described below. Each tree attempts to independently classify a given input. A RF uses multiple trees to reduce the variance and bias of the model, through providing each tree with only a random subset of the available features. Each tree makes a choice between the available classes for the input using the information available to it, and the final output `probability' is the fraction of trees which decided on each class. RFs are a supervised method requiring a training set, and are extremely versatile, assigning probabilities when classifying and so allowing ranking of candidates. RFs have been extensively discussed elsewhere, and we refer the reader to \citet{Richards:2011ji,Richards:2012ea,McCauliff:2015fb,Armstrong:2016br} and multiple others for detail of their methodology, tree classification methods and examples of their use in an astronomical context. Our \texttt{autovet} code is a wrapper for the RF built into the \texttt{scikit-learn} module \citep{Pedregosa:2011tv}. The code can be found at \footnote{https://github.com/DJArmstrong/autovet}.

\subsection{Features}
\label{sectfeatures}
A Random Forest uses features passed to it to perform classifications. These features could be any relevant data, such as planet radius or the $\theta_1$ statistic discussed above. RFs are typically robust to uninformative features, meaning several can be trialled when building a classifier. We extract features based on the signal detection, the transit and fits to it, the overall phase folded lightcurve and several diagnostics designed to spot instrumental artefacts. The features we implement are shown in Table \ref{tabfeatures}.

\begin{table*}
\caption{Features passed to the RF, ordered by feature importance calculated using the TEST18 data. A random feature was included for calibration, but not used in final classifications. Features in italics are dropped for the no-depth RF.
}
\label{tabfeatures}
\begin{tabular}{llr}
\hline
Feature & Importance & Description \\
\hline
SOM\_$\theta_1$ & 0.21593 & Transit shape statistic \citep{Armstrong:2017cp} \\
TransitSNR & 0.13951 & Transit depth over standard deviation of out of transit lightcurve \\
SOM\_Distance & 0.10820 & Distance of transit shape to nearest SOM pixel\\
Trapfit\_t14phase & 0.06803 & Transit duration 1st to 4th contact points from trapezoid fit, phase units\\
\textit{Trapfit\_depth} & 0.06652 & Depth of trapezoid fit, relative \\
Fit\_aovrstar & 0.06438 & Semi-major axis of transit fit, units of stellar radius \\
RANK & 0.05769 & Rank of BLS detection on that candidate \\
\textit{Even\_Trapfit\_depth} & 0.03752 & As Trapfit\_depth, even transits only \\
\textit{Fit\_rprstar} & 0.03560 & Planet to star radius ratio from transit fit \\
\textit{Odd\_Trapfit\_Depth} & 0.03385 & As Trapfit\_depth, odd transits only \\
GAP\_RATIO & 0.02847 & Fractional width of largest gap in phase-fold\\
NPTS\_TRANSIT & 0.02038 & Number of datapoints in transit\\
\textit{Even\_Fit\_rprstar} &0.01529 & As Fit\_rprstar, even transits only \\
DELTA\_CHISQ & 0.01502 & Difference in $\chi^2$ for BLS detection peak against baseline \\
\textit{Odd\_Fit\_rprstar} & 0.01236 & As Fit\_rprstar, odd transits only \\
NBOUND\_IN\_TRANS & 0.01054 & Number of night-boundaries within the transit box\\
AMP\_ELLIPSE & 0.00752 & Amplitude of ellipsoidal variation in the lightcurve\\
MaxSecDepth & 0.00671 & Maximum depth of a secondary eclipse\\
ntransits & 0.00589 & Number of observed transits\\
MaxSecSig & 0.00567& Maximum secondary eclipse significance\\ 
SN\_ELLIPSE & 0.00513 & SNR of ellipsoidal variation\\
SN\_ANTI & 0.00503 & SNR of strongest inverse transit detection\\
Full\_partial\_tdurratio & 0.00429 & Ratio of Trapfit t14phase to Trapfit t23phase\\
SDE & 0.00407 & BLS detection SDE\\
PointDensity\_ingress & 0.00373 & Density of points in ingress relative to average density in lightcurve\\
PointDensity\_transit & 0.00345 &  Density of points in transit relative to average point density of lightcurve\\
Even\_Odd\_trapdurratio & 0.00265 & Duration ratio (1st to 4th contact) between trapezoid fits to even and odd transits\\
Scatter\_transit & 0.00201 & Standard deviation of points in transit relative to that of lightcurve \\
Even\_Full\_partial\_tdurratio & 0.00189 & Full\_partial\_tdurratio for even transits only\\
MaxSecSelfSig & 0.00181 & Maximum secondary eclipse significance, normalised by other secondary detections\\
Even\_Odd\_depthdiff\_fractional & 0.00180 & Fractional depth difference between odd and even transit fits\\
Even\_Odd\_trapdepthratio & 0.00178 & Depth ratio between trapezoid fits to even and odd transits\\
Odd\_Full\_partial\_tdurratio & 0.00132 & Full\_partial\_tdurratio for odd transits only\\
\textit{Fit\_depthSNR} & 0.00116 & Depth from transit fit normalised by fit error \\
Fit\_chisq & 0.00115 & Best fitting $\chi^2$ of transit model \\
\textit{Odd\_Fit\_depthSNR} &  0.00096 & Fit\_depthSNR for odd transits only \\
missingDataFlag & 0.00091 & Fraction of missing data within 2.5 transit durations of transit\\
Even\_Odd\_depthratio & 0.00087 & Depth ratio between fits to even and odd transits\\
\textit{Even\_Fit\_depthSNR} & 0.00087 & Fit\_depthSNR for even transits only \\
\hline
Random & 0.00066 & Random feature\\
MaxSecPhase & 0.00063 & Phase of maximum detected secondary eclipse\\
\hline
\end{tabular}
\end{table*}

Initially we included several additional features incorporating information on noise inherent in the lightcurve: the root-mean-square, point-to-point percentiles, and median average deviation for example. During development it became clear that these features were surprisingly unhelpful; firstly, they only convey information indirectly, in that higher lightcurve noise only means a reduced chance of a planet for some transit and planet configurations. Secondly, as our synthetic injections were only injected into some fields due to computing constraints, using lightcurve noise related features biased the results to candidate lightcurves matching the distribution of noise in those fields. Fainter stars, more blended fields, and to an extent even poorly represented telescopes became downweighted, in a way that was hard to recover. We found that the simplest method to avoid these issues was to remove such features entirely. We show the effect this has on classifier performance in Section \ref{sectResults}.

There is a philosophical choice to make with regards to the features one makes available. One option is to incorporate knowledge of the known planet distribution; planets of a certain radius and orbital period are more likely than others, for example \citep{Fressin:2013df}. If trying to form a physically valid estimate of the likelihood that a given signal is a planet, then such an approach is likely necessary,  but would require an accurate knowledge of the planet distribution, and care as to how this distribution was reflected in the output from the survey and synthetic transits used to train the model. Incorporating this information is somewhat analogous to adding a prior to the model.

However, when performing a blind search, incorporating knowledge of the `known' planet distribution risks biasing the output rankings towards what you expect. This is particularly troublesome when considering new planet searches, as each survey has its own distribution of target stars and data characteristics, which will cause subtle or not-so-subtle effects in the observed planet distribution. Deliberately ignoring knowledge of some parameters, such as planetary radius, is then analogous to using an uninformative prior. The best approach is situation dependent and far from clear cut. Here in the context of a blind search using a new survey, we opt to exclude the planet radius and orbital period as direct parameters in an attempt to leave the resultant classifications as independent of prior knowledge as possible, and to avoid potential pitfalls resulting from incomplete information in our synthetic transit distribution. We do however include the directly fitted depth and duration of the transit signal, which while connected to the planet radius and orbital period provide strong constraints on eclipsing binaries.

Our injected transits are also limited by the parameter distributions used. In particular, the transit depths injected were between 0.1 and 2\%. Several planetary systems have transits larger than this depth, such as NGTS-1b \citep{Bayliss:2017bw}, and are as such likely to be downweighted. Including transit depth in the model is desirable as it is a strong predictor for eclipsing binaries. We ameliorate this issue by proceeding with two parallel models, one including transit depth related features (WD-RF) and one without (ND-RF), which will be each be tested. The features shown in italics in Table \ref{tabfeatures} are the ones removed in the ND-RF model. The TransitSNR feature is included in both models, despite being related to the transit depth, as we found it was integral to model performance. Leaving TransitSNR in still allowed deep transits to be recovered in the ND-RF model during testing.

\subsection{Transit Fitting}
\label{sectfitting}
Several of the features in Table \ref{tabfeatures} are derived from fits to the candidate signal, either to the whole lightcurve or to even or odd transits only. We perform two fits, one using a Mandel-Agol transit model \citep{Mandel:2002bb} implemented through the \texttt{batman} code \citep{Kreidberg:2015di} and one using a trapezoid. 

\noindent\textbf{Transit fit}: Free parameters are the epoch, orbital period, semi-major axis over the stellar radius, and planet-star radius ratio. We fix the inclination at $\frac{\pi}{2}$, the eccentricity at 0, and limb darkening to a quadratic law with parameters [0.1,0.3]. This is an intentionally oversimplified fit; our aim is to obtain approximations to what the parameters would be if the candidate was planetary in origin, even for extremely noisy signals which will often not be planetary or indeed astrophysical. As the method is intended to be used automatically on large datasets, no fits will be checked by eye, and hence only the most integral parameters can be fit for. The fit is performed using a least squares minimisation for simplicity. All features in Table \ref{tabfeatures} containing `Fit\_' derive from this method.

\noindent\textbf{Trapezoid fit}: Free parameters are the phase of transit, depth, time between 1st and 4th contact, and time between 2nd and 3rd contact. The orbital period is fixed. The fit is performed using a least squares minimisation for simplicity. All features in Table \ref{tabfeatures} containing `Trapfit\_' derive from this method.

\subsection{Feature Importance}
\label{sectfeatimp}
RFs allow estimation of the features contributing the most to the decision making process. Feature importance is estimated by considering where in the component Decision Trees features are considered - the higher up the tree, the more power a given feature has over the final classification. When averaged over the multiple Trees in a RF, this provides a ranking of features. Feature importances are shown in Table \ref{tabfeatures}, where features are ordered by their importance to highlight the most significant.

To provide calibration, a feature consisting of randomly generated numbers was added to the training set, and its importance evaluated. The one feature with an importance less than this random feature was excluded from the final classifier.

\subsection{Centroid Information}
Due to its extremely precise autoguiding, NGTS is the first ground based telescope to routinely use shifts in the centroid position of a target during transit as a vetting method, achieving precision of 0.75 milli-pixel on average, and 0.25 milli-pixel in the best cases. The full description of this process is described in \citet{Gunther:2017kd}. We trialled incorporating features marking significant centroid signals in the autovetter, but finally adopted a hybrid system whereby the centroid code independently flags certain targets. This is because certain forms of false positive, such as a blended transiting planet, might cause significant centroid shifts while still being interesting candidates that we may want to follow-up. Incorporating centroid related features in the RF would potentially downrank such candidates in a non-recoverable way. Furthermore, using centroid features would require simulating them for the synthetic transits, a somewhat arbitrary process involving significant human input in terms of the simulated distribution and thresholds used. When GAIA DR2 becomes available, it may be possible to incorporate centroid information in a sophisticated assessment of the probability of a given signal to lie on a given GAIA target, and this is planned in future development.

\subsection{Optimization/parameters}
\label{sectopt}
The most important parameters defining the RF structure are:
\begin{itemize}
\item $n_\textrm{est}$ The number of decision trees. In general, more trees represent improved classification at the cost of more computer resources, with diminishing returns.
\item $mf$ The maximum number of (randomly selected) features considered at each split within a tree. A typically good value is the square root of the number of features.
\item $d$ The maximum depth (number of splits) each tree can have
\item $ms$ The minimum number of samples required to split a node in the tree.
\end{itemize}

To optimize these parameters we use the TEST18 dataset. We perform a grid search over $n_\textrm{est}$ in $\left[100,300,500\right]$, $mf$ in $\left[2, 3, 5, 6, 7, 9\right]$, $d$ in $\left[2, 5, 8, 11, \textrm{None}\right]$, and $ms$ in $\left[2, 3, 4, 5\right]$. For each combination, the out-of-bag (OOB) score was extracted from the RF. Each tree in the forest is fit using a random sample of the training data. The OOB score is the accuracy of the classifier on each input using only trees which were not trained using that input, and provides a quick estimation of the accuracy of the RF. The best OOB score (of 96.4\%) was found for $n_\textrm{est}=300$,$mf=5$, no max depth and $ms=3$. We adopt these values for $mf$ and $ms$, but apply a restriction on max depth, setting $d=8$. This is because although apparently better results are found for no limit on tree depth, a RF with no max depth is prone to overfitting, producing confident results in regions of parameter space not supported by the data. A larger depth supports more parameter space complexity, which is not always justified. The difference in OOB score is marginal, at 95.5\%.

As this classifier forms part of the NGTS pipeline, we put further effort into minimising the processing time required. With the above fixed parameters, we varied $n_\textrm{est}$, finding that significant gains in OOB score stop being made after an $n_\textrm{est}$ of 200. As such, we use 200 for our models. We do not expect this optimisation to change significantly through different pipeline versions, as while different versions will change the data quantity and specific detrending characteristics, the nature of the classification problem itself will not change. 

\section{Results}
\label{sectResults}
\subsection{Classification Metrics}

With the form and parameters of the RF fixed, we turn to the CYCLE1706 dataset for testing. We retrain the two models with CYCLE1706, and use cross-validation to test their performance. Cross-validating consists of successively excluding blocks of inputs from the classifier, with the inputs in a random order. The classifier is then trained on the non-excluded data, and the trained classifier used to classify the initially excluded inputs. A new classifier is trained for each excluded block. In this way, classifications are obtained for each input using classifiers which were not trained on that input. The NGTS pipeline will continue to develop, and hence using CYCLE1706, a version which the RFs were not optimised on, is the ideal test for how they will perform in practice. It is necessary to retrain the models however, now and with each new pipeline version, to account for the varying dataspan and noise properties which different versions present.

Classification problems are typically measured using a series of standard metrics. Here we use three: the Precision (the fraction of signals that are classified as planets which are injected planets), the Recall (fraction of injected planets which are classified as planets), and the AUC (area under the receiver-operator characteristic curve, see \citealt{Fawcett:2006gr}). These results are shown in Table \ref{tabscores}, for models with and without lightcurve noise related features (see Section \ref{sectfeatures}) and with and without transit depth related features (see Section \ref{sectfeatures} and Table \ref{tabfeatures}). The overall performance of the classifier is largely unaffected by these choices, showing a slight improvement as more features are added, but this masks divergent performance on specific subsets of candidates. In summary, noise related features introduce biases associated with specific fields and cameras used to generate the training set, and are hence removed from here on. Depth related features, when included, downrank several known planets with deep transits ($>2$\%), as this is the limit of our synthetic injections. Removing these features recovers these planets, at the expense of greater weakness to eclipsing binaries. For the remainder of the paper, we explore models without noise related features, but with (WD-RF) and without (ND-RF) depth related features, as scientifically interesting planets such as NGTS-1b \citep{Bayliss:2017bw} can present deep transits.

For the WD-RF model, we show scores for both cross validation and for performance on an unseen test set, which consisted of one field of CYCLE1706 data not seen by the classifier. The test set consisted of 2251 synthetic injected planets and 348 real candidates, after period cuts were run as described in Section \ref{secttrainingset}. The numbers are unbalanced as we used a single NGTS field with a limited number of candidates, leading to an anomalously high precision but otherwise comparable results. We proceed using the cross-validation results, as they are similar if slightly lower and incorporate more of the available data to produce the values.

We note that the scores are not precisely correct, as their calculation assumes each training set is completely accurate. As some fraction (unknown as yet) of our real candidates will be real planets, this assumption does not hold. Given known planet occurrence rates and the hit rate of other ground based surveys we expect the proportion of real planets to be low enough to make little difference to the scores. In \citet{Guenther:2017a}, we previously estimated that $\sim$97\% of all initial BLS detections in NGTS are caused by false positives (eclipsing binaries and background eclipsing binaries) and false alarms (systematic noise). This value is in line with the findings from other surveys. In comparison, initial detections in CoRoT and HAT contained 95-98\% of false signals \citep{Almenara:2009fb, Latham:2009bx, Hartman:2011}.

\begin{table}
\caption{Cross-validated classification scores for Random Forest models.}
\label{tabscores}
\begin{tabular}{lllr}
\hline
Features used & Precision & Recall & AUC \\
\hline
\textbf{No Depth, No Noise} & 0.892 & 0.905 & 0.976\\
\textbf{With Depth, No Noise} & 0.897 & 0.910 & 0.978 \\
\hspace{3mm}\textbf{\textit{Cross-Validation}} & 0.897 & 0.910 & 0.978 \\
\hspace{3mm}\textbf{\textit{Train-Test}} & 0.99 & 0.93 & 0.979 \\
No Depth, With Noise &  0.896& 0.910 & 0.979 \\
With Depth, With Noise & 0.901  & 0.914 & 0.980 \\
\hline
\multicolumn{4}{l}{Precision and recall calculated using a threshold of 0.5}\\
\end{tabular}
\end{table}

\begin{figure}
\resizebox{\hsize}{!}{\includegraphics{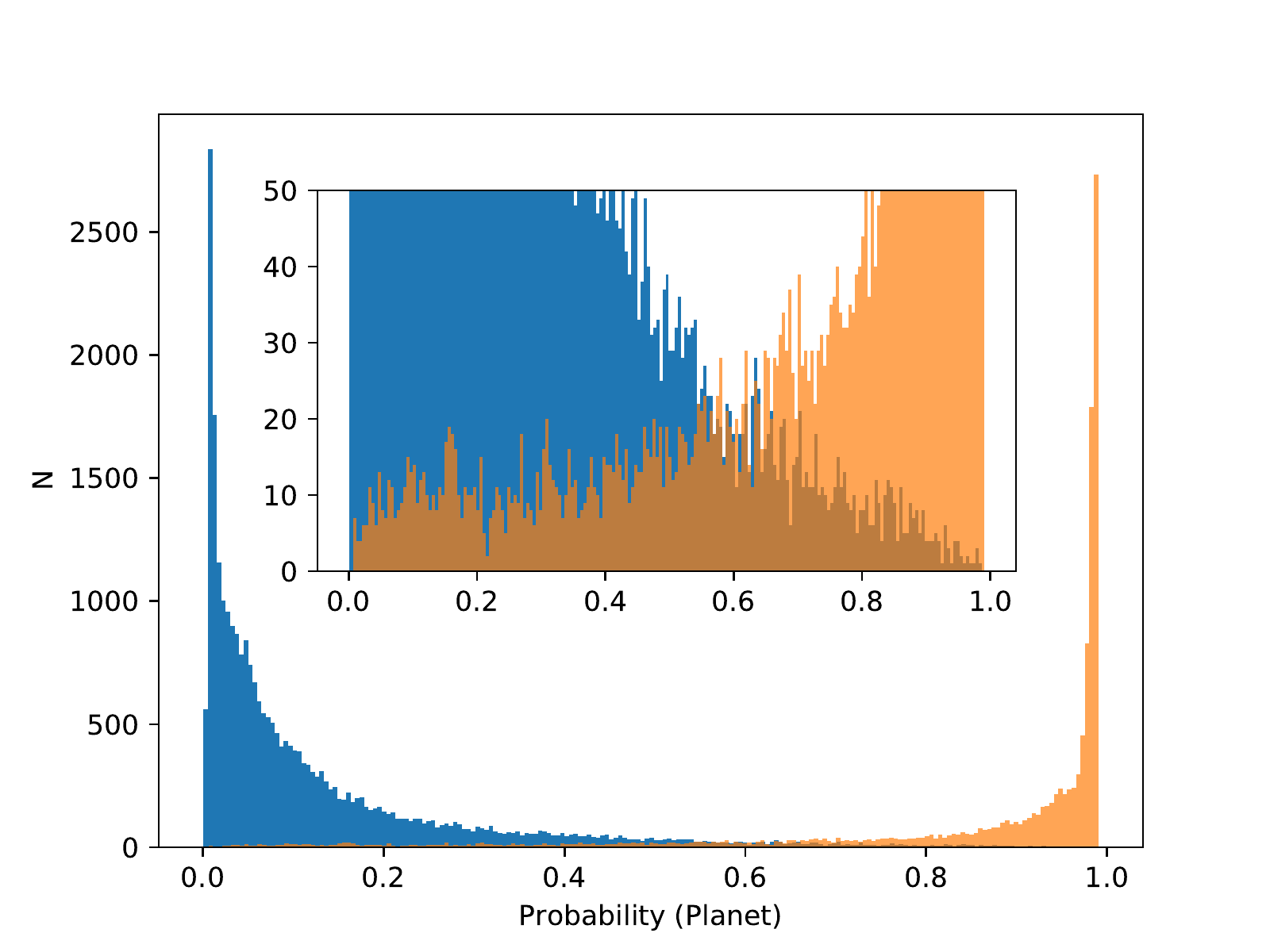}}
\caption{RF output planet probabilities for the real candidates (blue) and synthetic injected transits (orange). A zoomed in inset shows the few misclassified injections. Real candidates with high planet probabilities are strong candidates for further followup, unlike the majority of signals.}
\label{figcvprobs}
\end{figure}

The distribution of probabilities output by the WD-RF classifier is shown in Figure \ref{figcvprobs}, for the real candidates and synthetic injections. Probabilities are calculated using cross-validation, in which the candidates are randomly ordered, and successive blocks of 500 are removed. For each removed block, the classifier is trained on the remainder of the candidates, and then used to classify the withheld 500. The process is repeated with the classifier retrained each time until all candidates have been classified. The efficacy of the classifier at separating the two classes is evident. A tail of real candidates with high probabilities of being in the 'Planet' class can be seen, and these represent potential planets discovered by the classifier. We note that there are two use cases for the classifier here: one to identify high-priority candidates, and the other to rank all the candidates. In the context of ranking, even a score of 0.2 puts a planet above \mytilde80\% of the signals.

\subsection{Key Features}

Feature importances are calculated as described in Section \ref{sectfeatimp}. We list the top 10 features here for the ND-RF model, with a more detailed description than given in Table \ref{tabfeatures}. Note that the feature importances listed in Table \ref{tabfeatures} are for the WD-RF model, which changes the order slightly.

\begin{figure}
\resizebox{\hsize}{!}{\includegraphics{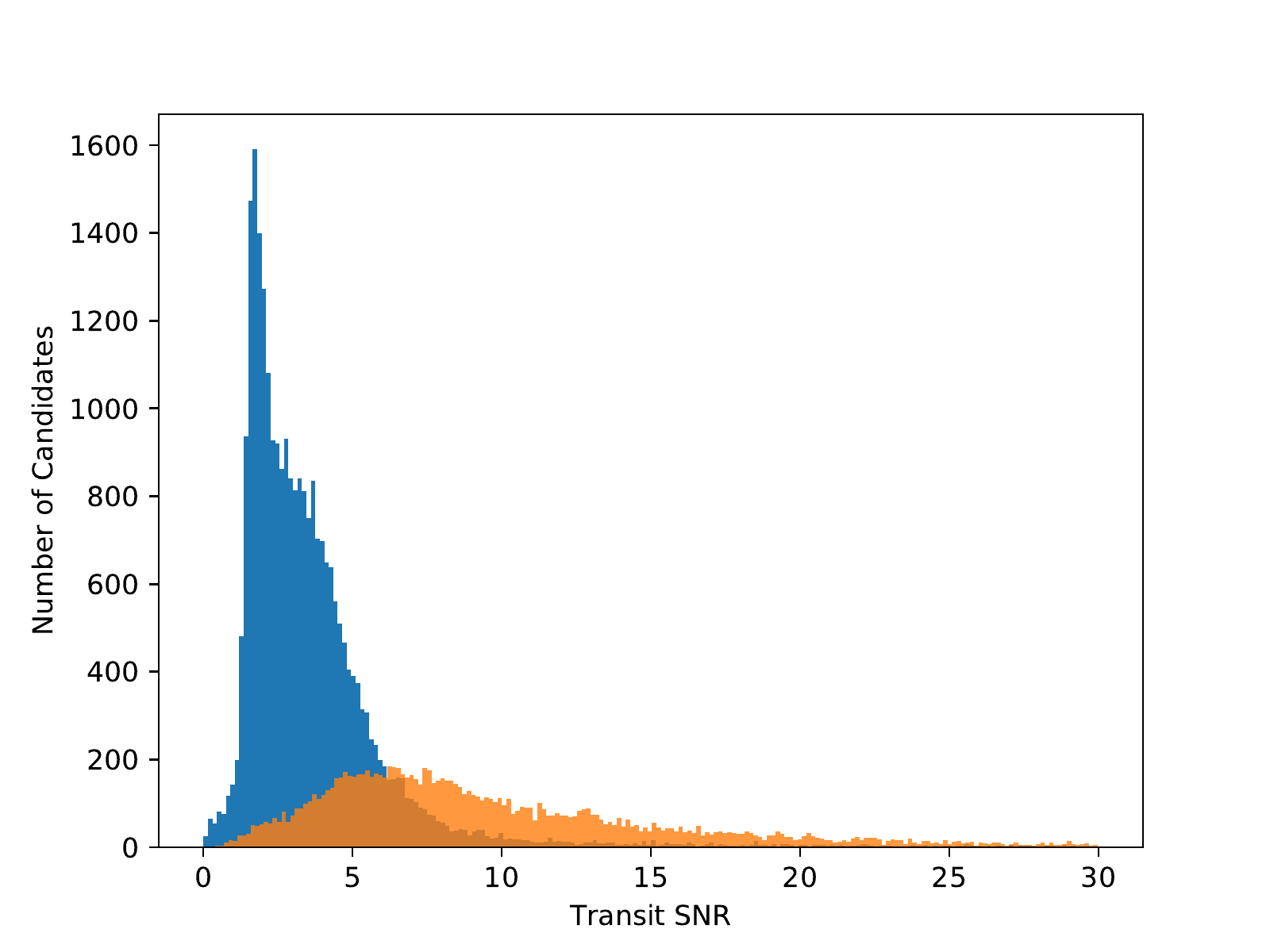}}
\caption{As Figure \ref{figtheta1dists} for the TransitSNR feature.
}
\label{figtsnrdists}
\end{figure}

\begin{figure}
\resizebox{\hsize}{!}{\includegraphics{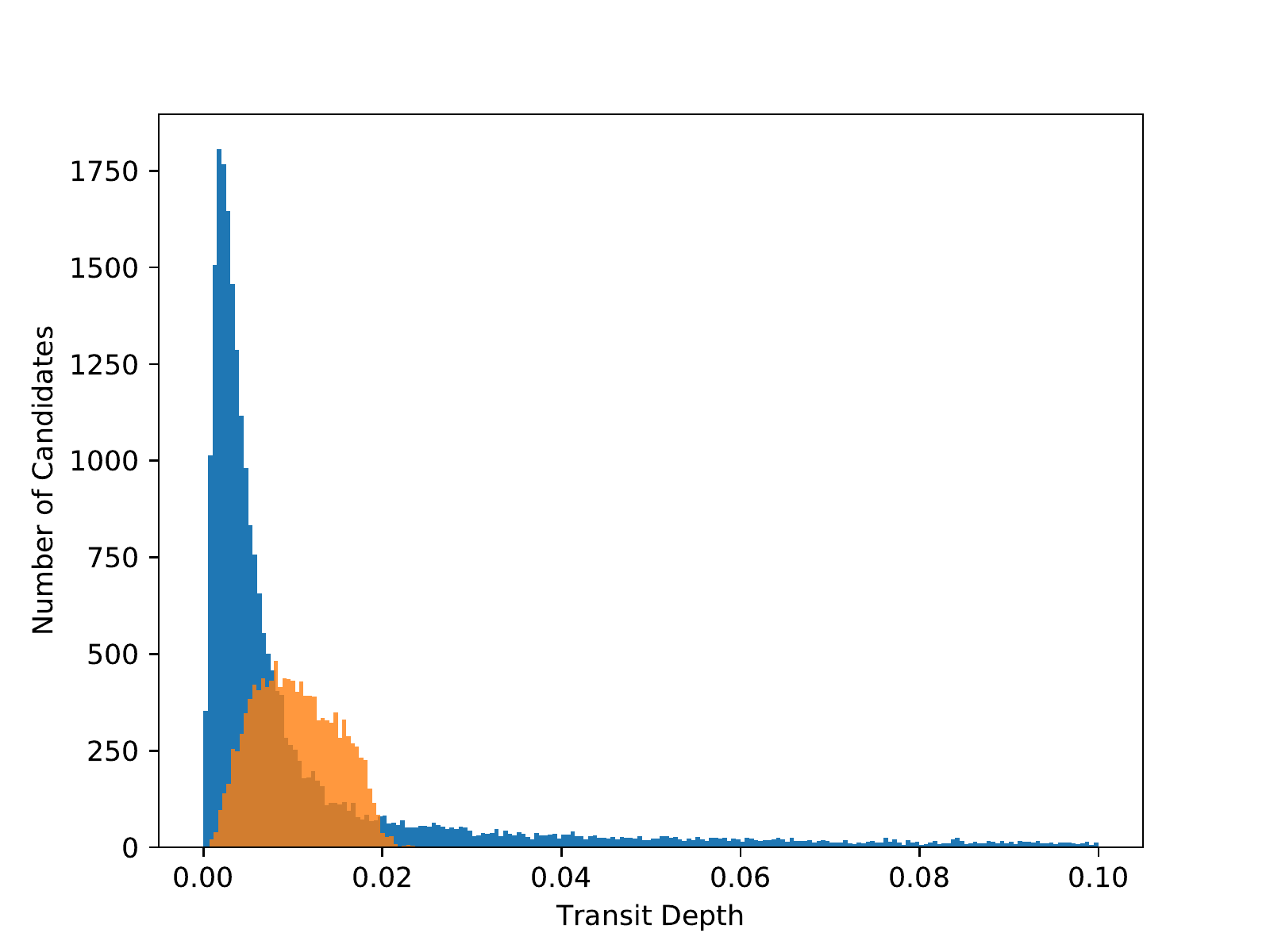}}
\caption{As Figure \ref{figtheta1dists} for the Trapfit\_depth feature. The histogram for real candidates (blue) extends to higher depths which are not shown for clarity.
}
\label{figtddists}
\end{figure}

\begin{figure}
\resizebox{\hsize}{!}{\includegraphics{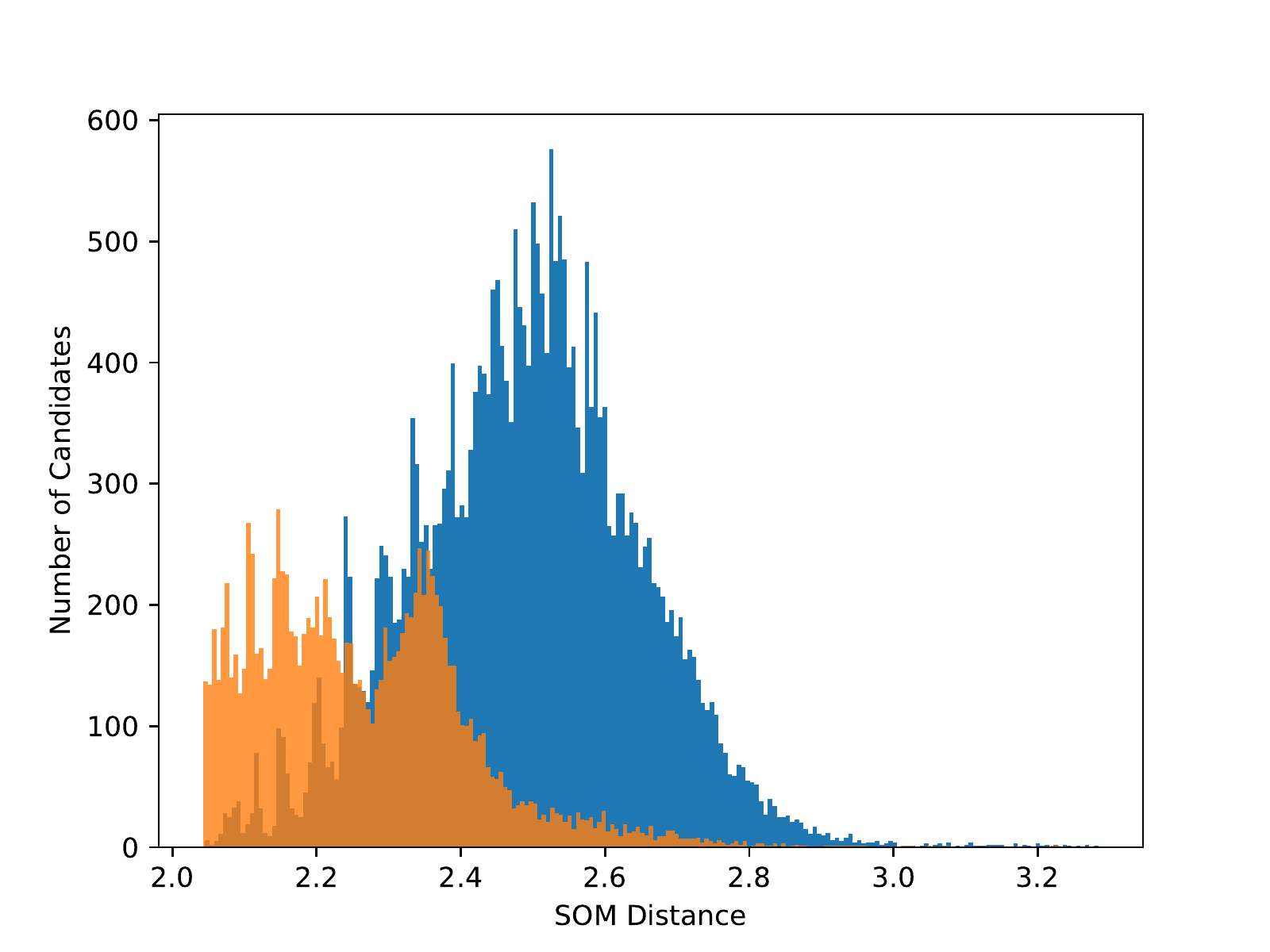}}
\caption{As Figure \ref{figtheta1dists} for the SOM Distance feature.
}
\label{figsddists}
\end{figure}

\begin{itemize}
\item \textbf{SOM\_$\theta_1$} 
Transit shape statistic, falling between 0 (for transits similar in shape to the set of false positives) and 1 (for transits similar in shape to the synthetic injections). The statistic is produced by comparing the binned transit to a SOM generated from the complete set of candidates and synthetic transits, and identifying the closest match. See \citet{Armstrong:2017cp} for a full description, and Figure \ref{figtheta1dists} for the calculated distributions.
\item \textbf{TransitSNR}
The signal-to-noise of the detected transit signal, measured on the transit duration. The phase-folded lightcurve is binned on 80\% of the transit duration (to avoid ingress and egress). TransitSNR is the bin value of the in transit bin divided by the standard deviation of the out of transit bins. Note that this measures the `detectability' of the transit. While the depth derived will not be accurate for highly V-shaped transits, TransitSNR provides a measure of how easily such transits will be seen by our BLS implementation. The distributions of TransitSNR are shown in Figure \ref{figtsnrdists}. We additionally show the distribution of transit depths as measured by the trapezoid fit in Figure \ref{figtddists} to show the difference between SNR and depth.
\item \textbf{SOM\_Distance}
When a signal is compared to the SOM and finds its closest match, the distance between the signal and this match is calculated. This is typically higher for false positives (as these have a wider range of possible shapes), and so the distance can help to resolve the classification in cases marginal in SOM $\theta_1$. The distribution of SOM Distance is shown in Figure \ref{figsddists}.
\item \textbf{Trapfit\_t14phase}
A trapezoid model is fitted to the detected signal (see Section \ref{sectfitting}). One of the outputs of this fit is the transit duration, in phase, between the 1st and 4th contact points.
\item \textbf{Fit\_aovrstar}
We also use \texttt{batman} \citep{Kreidberg:2015di} to fit a Mandel-Agol transit model to the detected signal (see Section \ref{sectfitting}). One output from this fit is the putative planet's semi-major axis in units of the stellar radius.
\item \textbf{Rank}
Our BLS algorithm outputs the top 5 periodic signals for each candidate. The rank (an integer between 1 for the strongest and 5 for the weakest) represents the order for the signal under question.
\item \textbf{GAP\_RATIO}
The fractional width of the largest gap in coverage in the phase-folded lightcurve.
\item \textbf{DELTA\_CHISQ}
The difference in $\chi^2$ between a box model and a flat model for the peak BLS detection.
\item \textbf{MaxSecDepth}
The phase folded lightcurve is scanned for secondary eclipses, between phase 0.3 and 0.7, using a box of width the transit duration. The most significant possible secondary eclipse in this region is found and the mean flux value in the eclipse extracted.
\item \textbf{AMP\_ELLIPSE} 
The amplitude of any ellipsoidal variation in the phase-folded lightcurve.
\item \textbf{NPTS\_TRANSIT} 
The number of data-points within the best fitting transit box.
\end{itemize}

\subsection{Known Planets}
\label{sectknownplanets}
Four published planets, 3 as yet unnamed confirmed NGTS discoveries and 42 previously flagged planetary candidates are available in the dataset for testing. These form ideal test cases for the algorithm. Results for the confirmed planets are shown in Table \ref{tabplanets}. For the ND-RF, all but one planet are detected with scores greater than 0.5. From a ranking perspective, we note that even the lowest scoring planet, with a score of 0.27, ranks in the top 15\% of candidate signals, demonstrating the efficacy of the RF in aiding detection. For the WD-RF, the planets with transit depth larger than 2\% are predictably downranked, in particular HATS-43b with a depth of 2.9\%.

\begin{table}
\caption{Scores for observed known planets}
\label{tabplanets}
\begin{tabular}{lllr}
\hline
Planet & ND-RF & WD-RF & Discovery Reference\\
\hline
HATS-43b &  0.95 & 0.04 & \citet{Brahm:2017uc} \\
WASP-98b &  0.51 & 0.11& \citet{Hellier:2014dqa} \\
WASP-68b & 0.86  & 0.93 &  \citet{Delrez:2014iv}\\
NGTS-1b &  0.73 & 0.65 & \citet{Bayliss:2017bw} \\
NGTS-xb &  0.27 & 0.39 & in preparation \\
NGTS-xb &  0.63 & 0.20 & in preparation \\
NGTS-xb &  0.93 & 0.95 & in preparation \\
\hline
\multicolumn{4}{l}{ND-RF=No depth features, WD-RF=With depth features}\\
\end{tabular}
\end{table}

\begin{figure}
\resizebox{\hsize}{!}{\includegraphics{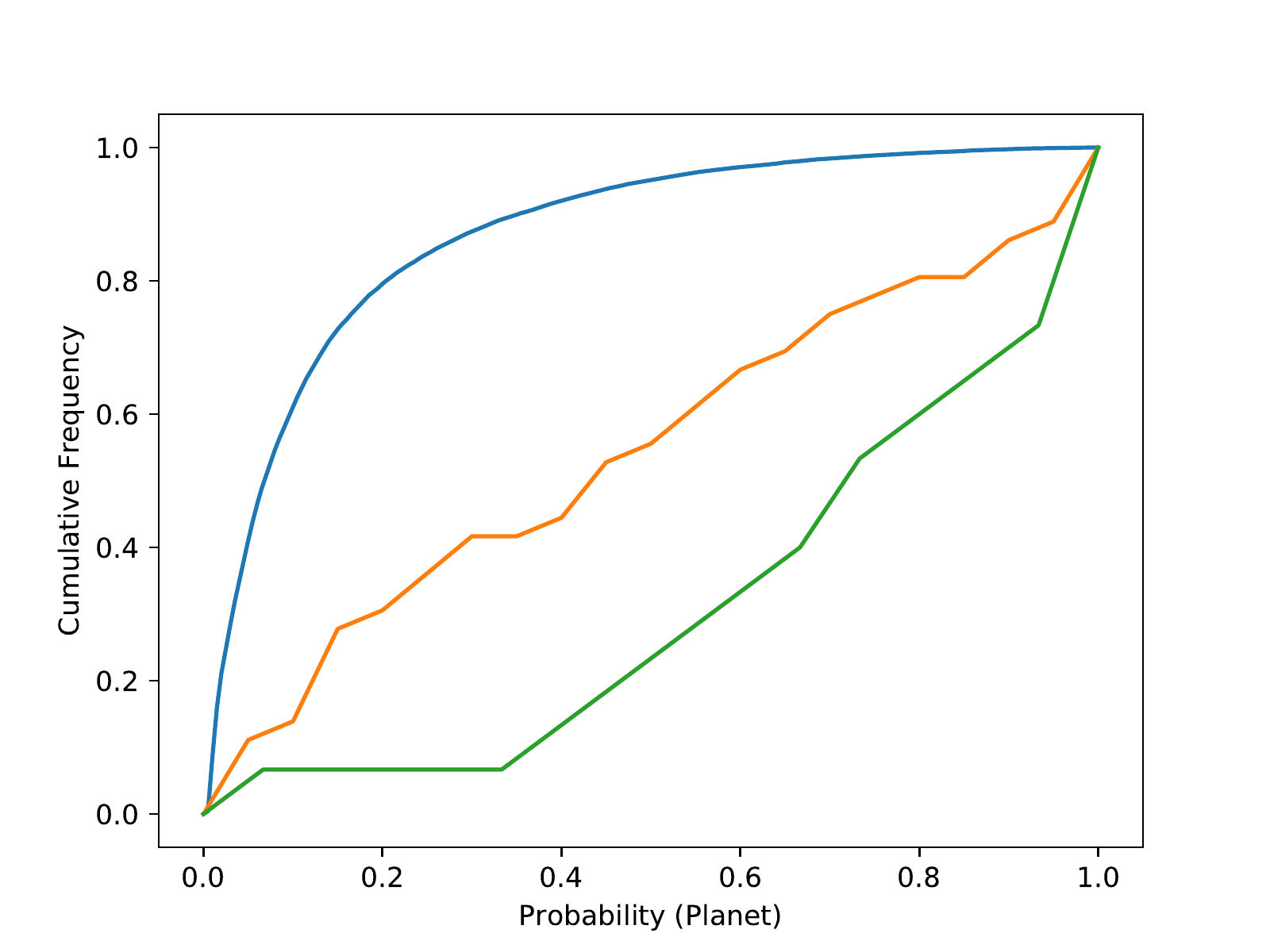}}
\caption{Normalised cumulative frequency of candidates for the total set of BLS signals, i.e. in the majority false positives (blue), for candidates already pre-selected by eye for follow-up (orange) and for the pre-selected candidates ignoring those deeper than 2\% or V-shaped (green). The difference of the orange and green curves to the blue shows the model's effectiveness at ranking good candidates higher than the background of signals.}
\label{figcumfreqs}
\end{figure}

The 42 flagged planetary candidates will contain some real planets and some false positives. The cumulative frequency as a function of their RF output for these candidates is shown in Figure \ref{figcumfreqs}, along with that for the entire set of signals studied. It is clear that the RF ranks the 42 candidates much higher than the general background of signals, and more so for a subset of the candidates. We note that there is a bias in the sample studied here - as these are early results from the survey, and pre-NGTS planets arise from ground based surveys, the sample of candidates and planets is typically larger, with deeper transits than we expect on average. Several of the highlighted candidates are shallower (in the Neptune radius regime) however, with promising follow-up work ongoing.

\subsection{Highly Ranked Candidates}

Both models highlight numerous candidates for further observations. The ND-RF flags 53 candidates with planet probability$>$0.9, and 1355 with probability$>$0.5, out of a total of 27498 that passed the period matching cuts. The WD-RF flags 50 and 1294 respectively. These are in the majority but not exclusively the strongest BLS signals detected on an object.

We cannot lay out all of these candidates here, but focus on the top 10 in each case to illustrate the power of the method. Five of these top 10 overlap between the ND-RF and WD-RF, demonstrating the robustness of the models. In the WD-RF, the top 10 contain 4 candidates now undergoing follow-up spectroscopic observations (1 newly flagged by this method), 3 candidates that were previously selected by eye which were revealed as double lined spectroscopic binaries from spectroscopy (none show obvious characteristics which would exclude them as planets from the lightcurve alone, demonstrating the importance of follow-up observations at this stage of development), 2 candidates which show clear interesting potentially planetary signals but which are blended with other targets in the photometric aperture, and one candidate which shows an interesting signal but which is likely a giant host star. These results are encouraging, with 7 out of 10 candidates worthy of further follow-up and the remainder understandable given the model inputs. These latter 3 cases demonstrate a weakness - at present no information is given regarding blending or host star type, and we plan to incorporate features with this information in future iterations. Such information will become readily available with the GAIA DR2 data release.

The additional 5 candidates highlighted by the ND-RF model show 2 candidates amenable to follow-up observations (one new, one previously scheduled), 2 strong but too deep candidates (4\% and 3\% transits, including one which is also blended in the aperture), and one interesting signal but on a likely giant host star.

\section{Validation Tests}
\label{sectVal}
\subsection{Sensitivity}

An advantage of using a synthetic array of injected transits is that we can test the performance of the RF on a known dataset, investigating our sensitivity and providing some interpretation of how the RF is working. We note however that such a study can only take us so far - the distribution of real planets in the data may be different, especially outside the range of the synthetic injections, and hence the RF may respond differently. Without a sample of detected planets from the mission this is impossible to test directly, but early results on detected planets are promising (Section \ref{sectknownplanets}).

\begin{figure}
\resizebox{\hsize}{!}{\includegraphics{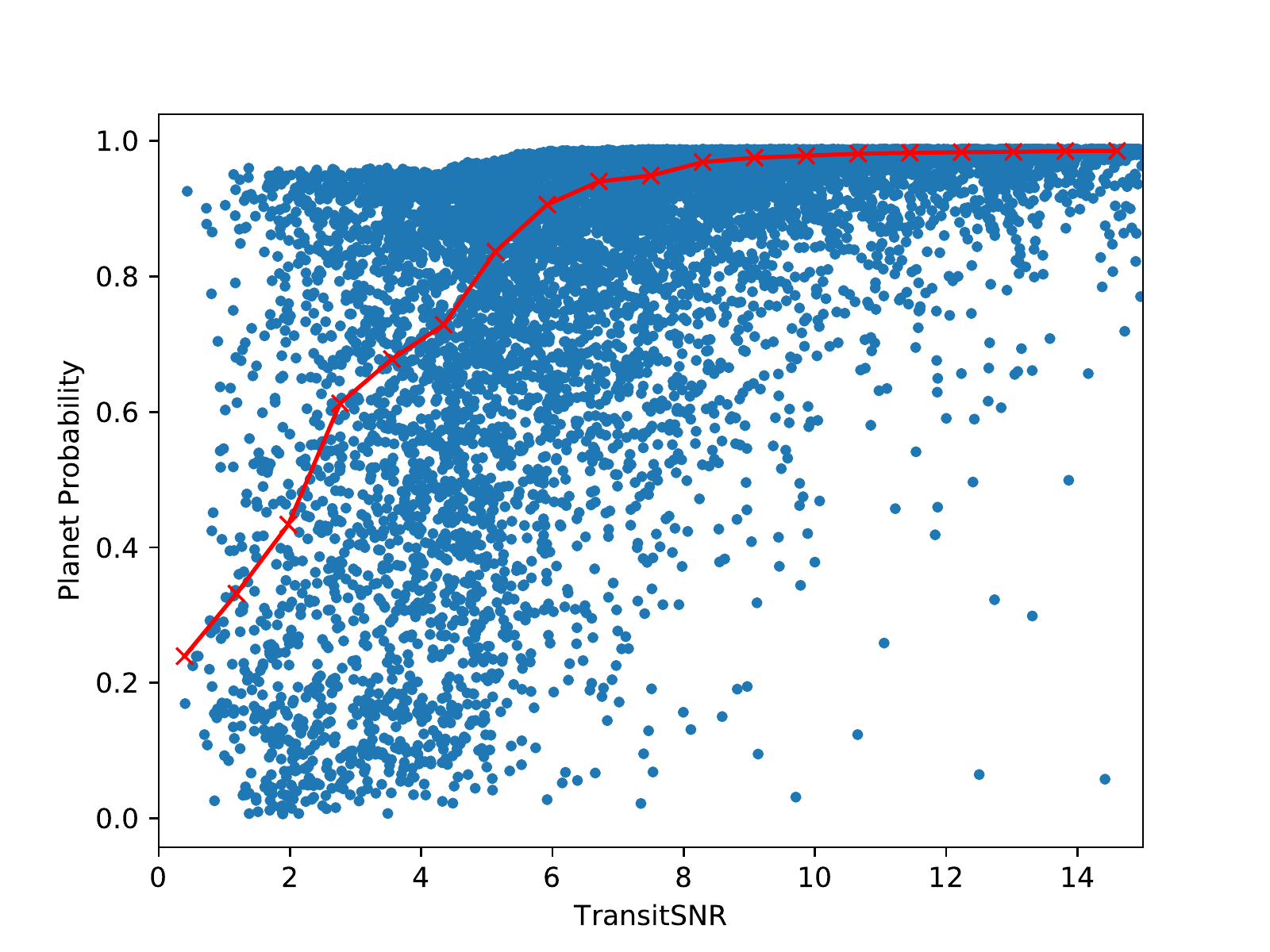}}
\caption{RF output planet probability as a function of transit SNR for the synthetic injected transits, with higher SNR cases not shown. The median planet probability for a series of bins is shown as red crosses.}
\label{figsnrvspprob}
\end{figure}

\begin{figure}
\resizebox{\hsize}{!}{\includegraphics{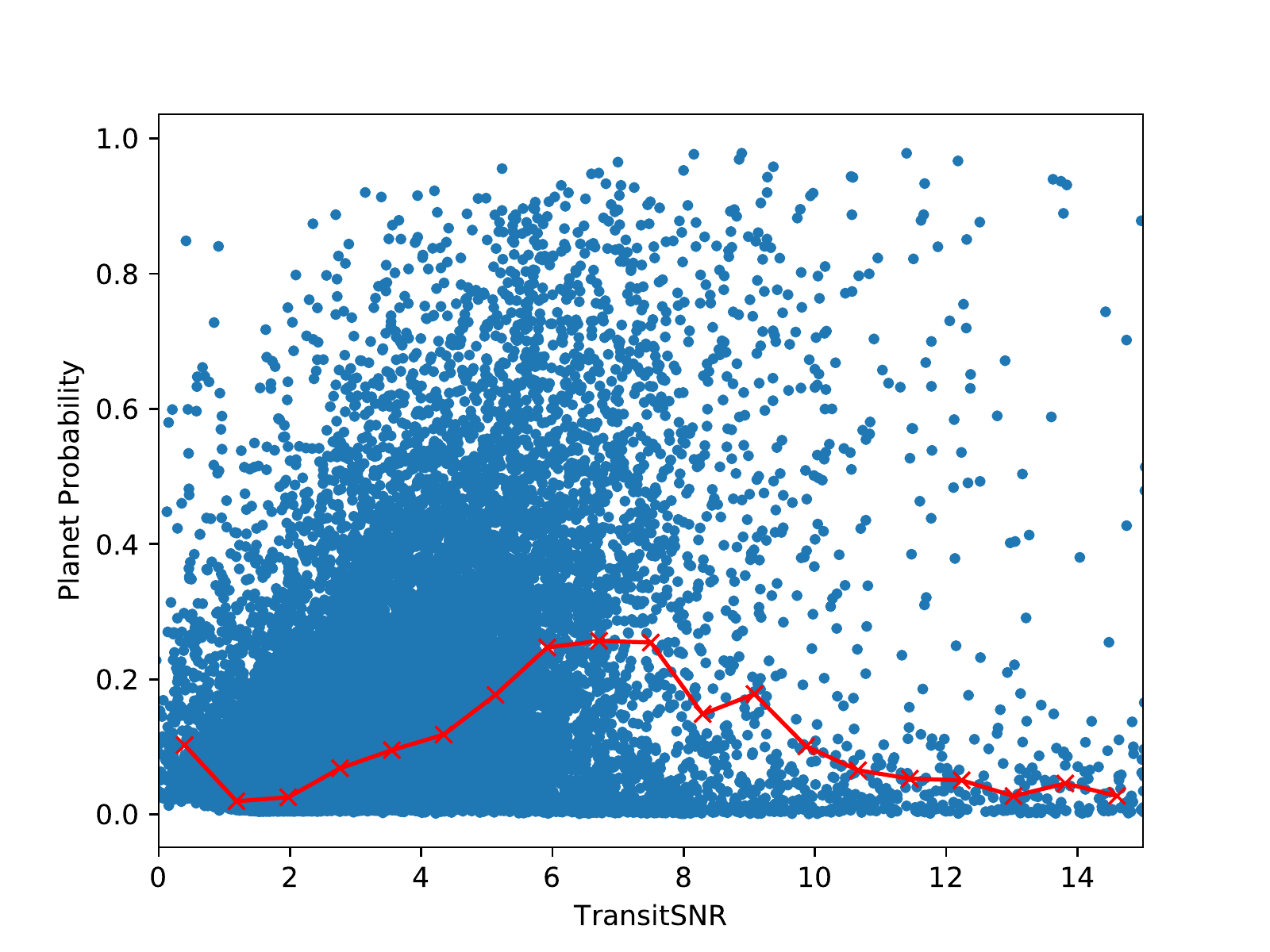}}
\caption{As Figure \ref{figsnrvspprob} for real candidates.}
\label{figsnrvspprob_cands}
\end{figure}

We begin by studying the parameters of the synthetic transits which were not recovered, using cross-validated probabilities in all cases and considering the WD-RF model. Figures \ref{figsnrvspprob} and \ref{figsnrvspprob_cands} show the output planet probability of the transits as a function of their SNR, for synthetic injected transits and for real candidates respectively. The SNR shown here is measured as the depth of the binned transit, normalised by the standard deviation of out-of-transit bins of the same duration. The data shows that our sensitivity begins to drop at SNR of 7-8, interestingly in line with the threshold of 7.1 used by the \textit{Kepler} pipeline \citep[][and reference therein]{Thompson:2017wu}, although their calculation of SNR is different. Below this threshold the uncertainty on the model increases. While lower SNR planets are not always detected, the dispersion increases such that even at the lowest measured SNR some injected planets are detected at high confidence.

\begin{figure}
\resizebox{\hsize}{!}{\includegraphics{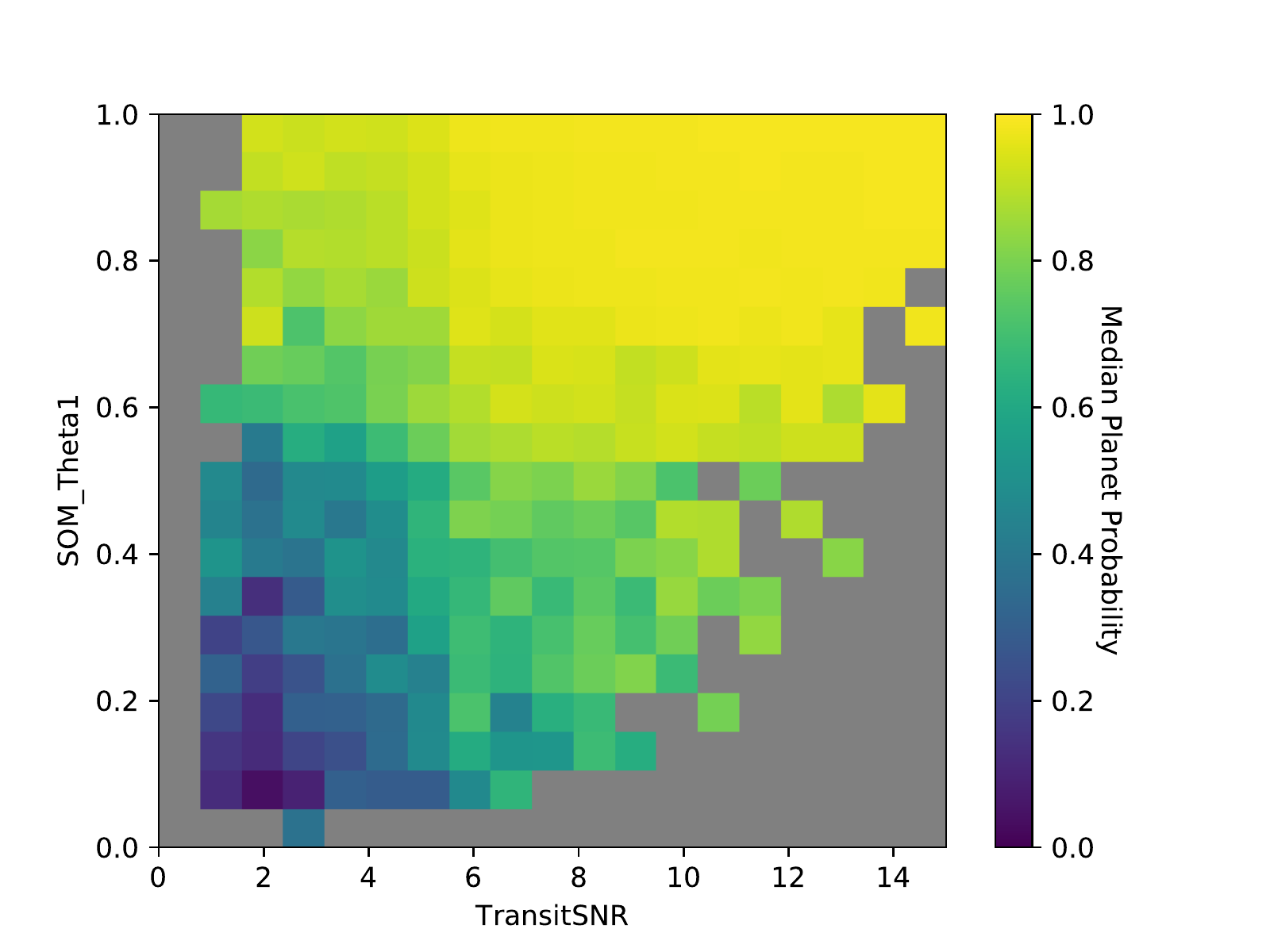}}
\caption{Median planet probability as a function of TransitSNR and SOM $\theta_1$ statistic, for the synthetic injected transits. Bins containing fewer than 5 candidates are in grey. Higher SNR cases are not shown.}
\label{figsnrvssom}
\end{figure}

\begin{figure}
\resizebox{\hsize}{!}{\includegraphics{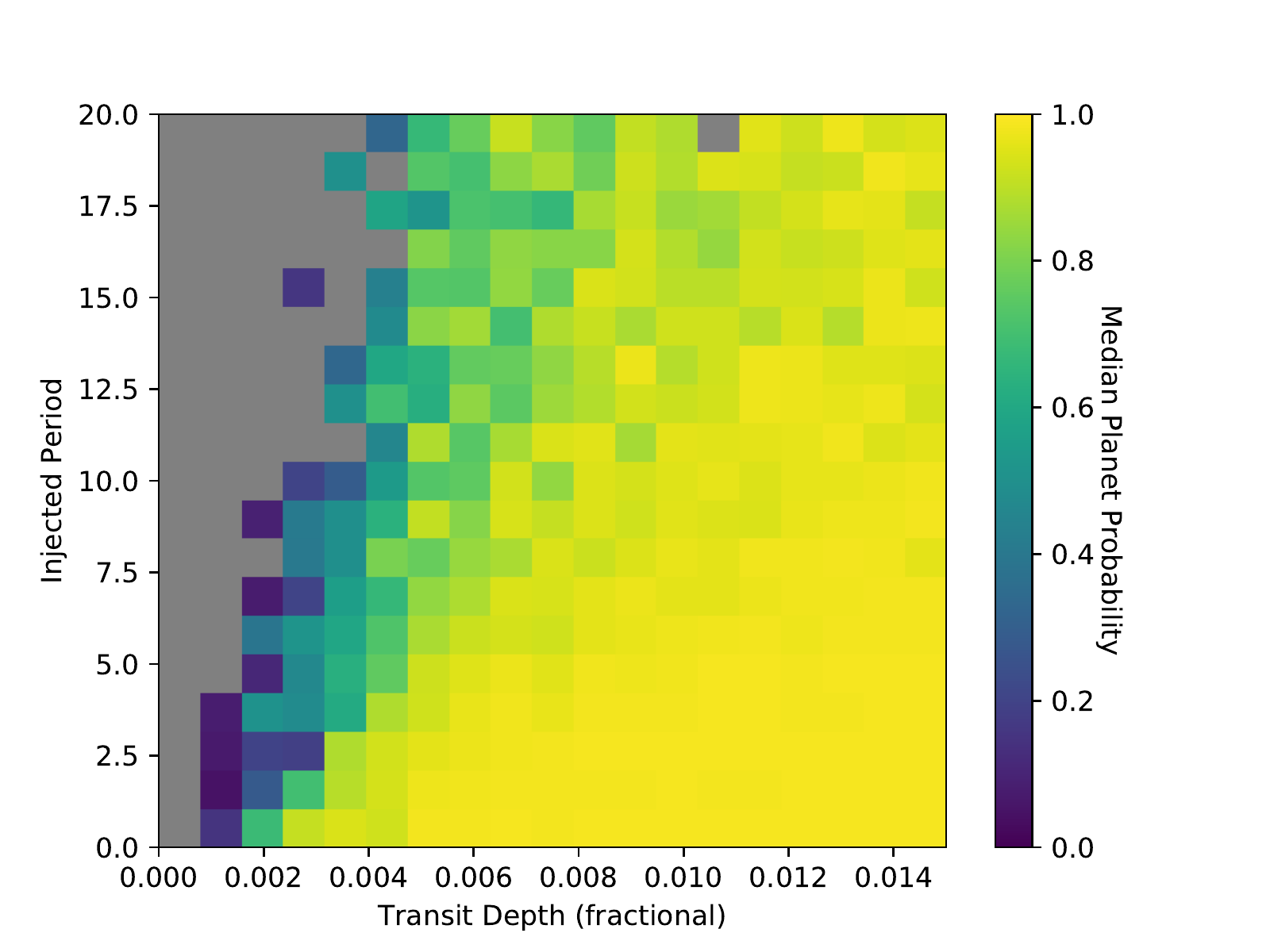}}
\caption{As Figure \ref{figsnrvssom} for injected transit depth and orbital period. Bins containing fewer than 5 candidates are in grey. Larger depth cases are not shown.}
\label{figdepthvsper}
\end{figure}

\begin{figure}
\resizebox{\hsize}{!}{\includegraphics{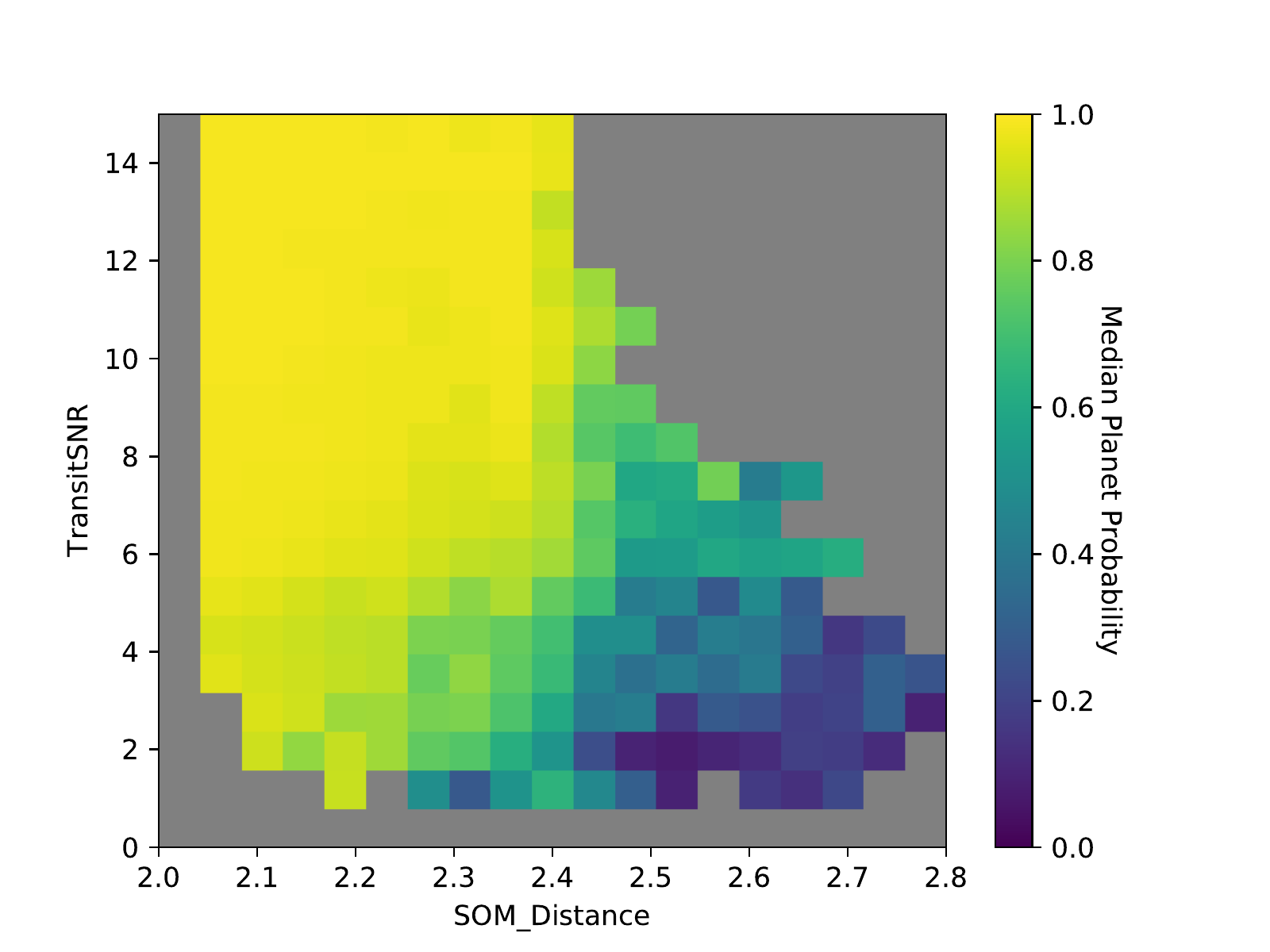}}
\caption{As Figure \ref{figsnrvssom} for TransitSNR and SOM\_Distance, for the synthetic injected transits. Bins containing fewer than 5 candidates are in grey. Higher SNR cases are not shown.}
\label{figsnrvssomdist}
\end{figure}

Having found the cut-off in sensitivity for SNR, we can investigate what it is that causes the RF to drop these planets. The most important feature used by the RF is the SOM $\theta_1$, and we plot the output planet probability as a function of this and the SNR in Figure \ref{figsnrvssom}. Low SOM $\theta_1$ and low SNR both cause a drop in planet probability. The cause of this is twofold; either low SNR transits are misinterpreted by the SOM as their shape is unclear, or grazing transits with a pronounced V-shape, and hence low $\theta_1$, have naturally lower depth and hence low SNR. Figure \ref{figdepthvsper} shows the output planet probability as a function of the transit depth and orbital period. Longer period and decreased depth are associated with lower planet probability as expected, as in each case the SNR will drop. However it is encouraging that for short periods where many transits are observed, there is little dependence on transit depth, showing that our models are not biased in this regard. The same is true for larger depths and longer periods. We note that as shown in Figure \ref{figsnrvspprob}, even the more extreme long period low depth transits are sometimes detected by the model. Finally Figure \ref{figsnrvssomdist} shows the effect of the SOM distance on synthetic transits. Transits only have a high distance when they are at low SNR and their shape is not well defined, highlighting that the criteria is not solely SNR but a combination of SNR and how the transit shape interacts with the noise in the lightcurve.

\subsection{Field Dependence}
\begin{figure*}
\resizebox{\hsize}{!}{\includegraphics{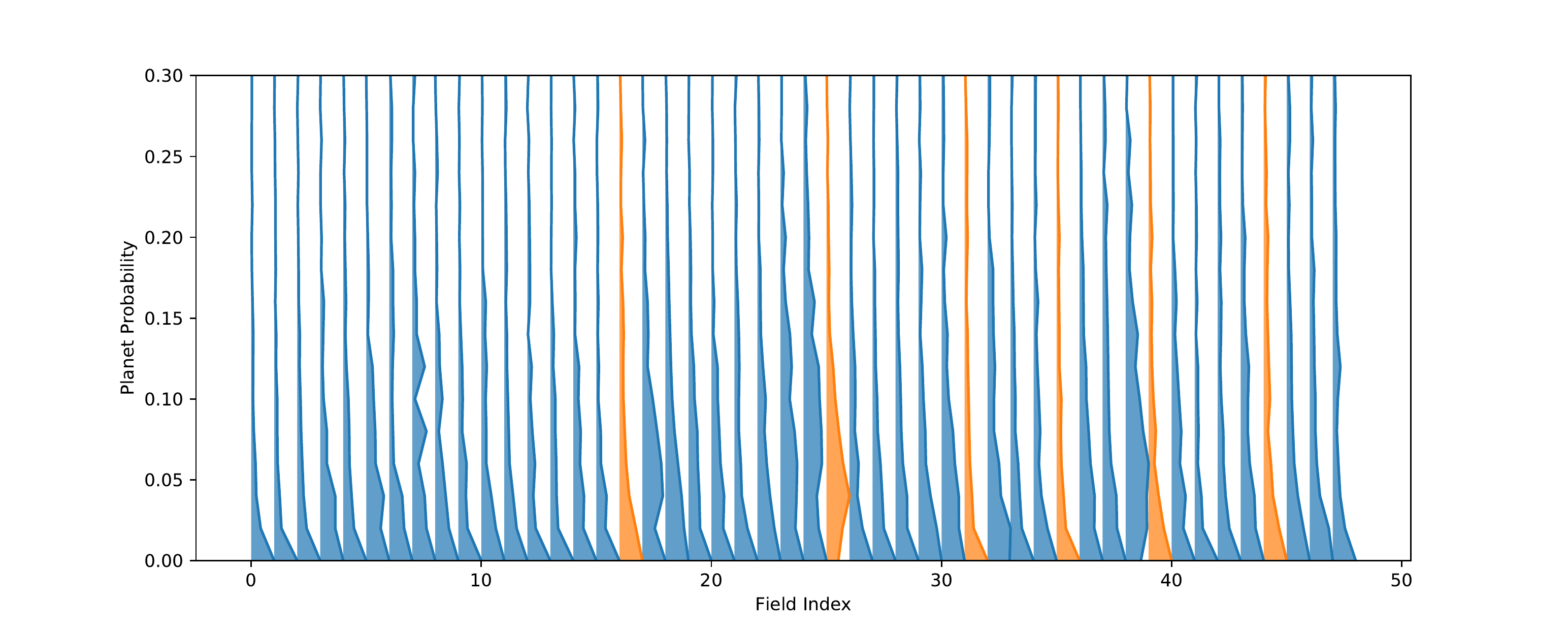}}
\caption{Planet probability distributions for the real candidates, as a function of observing field. Fields in orange are those where the synthetic transits were injected. Fields are ordered by median planet probability.}
\label{figfields}
\end{figure*}

As we injected transits into only a subset of observed fields, it is important to check that the RF is not biased towards candidates in these regions. This is potentially problematic, as fields are observed by different cameras, and can have differing noise properties, window functions, durations and crowding. We verify this by plotting the planet probability distribution for real candidates by field in Figure \ref{figfields}. If candidates from a field are downranked due to their membership in that field, the distribution from that field will be affected. The fields used for injections are highlighted, and are not divergent from the set.

\subsection{Outliers and Overfitting}

A common issue when applying machine learning techniques is overfitting; models can become confident about classifying regions of parameter space not supported by the data, especially when the model complexity is not justified by the quantity of training data. More problematically, if training data is sufficient but lacking in a particular region of parameter space, or outliers in the training set are present, problems can occur which may be harder to diagnose. This issue is particularly problematic where results on individual cases are important, as is the case in a survey searching for a relatively small number of planets. Overfitting may result in overconfident classifications, or erroneous outputs for specific objects. Such errors can be shown during testing, but if they occur in small regions of parameter space not covered by the test set then identifying the issue, or even that there was an issue, would be challenging.

Each machine learning method has options for guarding against overfitting. In the case of a RF, a key option is the maximum depth of the component decision trees. The depth is described in Section \ref{sectopt}, and is the number of splits each tree is allowed to make. Reducing the maximum depth guards against overfitting, as it reduces the potential complexity of the RF. For example, when optimising our models with no maximum depth imposed, we obtained tree depths of 25-30. Imposing a maximum depth of 8 did not significantly change the accuracy of the models, and as such we conservatively adopted this limit. Another option would be to search for outliers \citep[through for example an Isolation Forest,][]{Liu:kh} and remove these before training.

In this context, we have an injected synthetic distribution which by definition does not contain outliers (more accurately, any outliers in our specific parameter space arise because of peculiarities of the lightcurve rather than the planet itself, and hence will be rare). The real candidates will contain anomalous and unusual objects, but as these are classified as false positives in the training set they will only lead to the RF classifying other similarly unusual candidates as false positives, which in this context is desirable. Note that by anomalous here we describe unusual noise sources and combinations of features; such objects will not typically be scientifically interesting, and could in any case be targeted separately if desired. In essence, for the purposes of ranking candidates we do not consider outliers to be an issue here. However, should one need to trust the output planet probability for an individual candidate (for validation for example), the issue is much more important. In such a use case, subtle biases of the model must also be taken into account, and it is not yet clear how to adequately establish trust on individual inputs for often fundamentally uninterpretable machine learning methods. We leave that development for future work.

\section{Conclusion}
\label{sectConc}
We have presented a method for ranking candidates from a transiting planet search. Through incorporating random forests and self-organising-maps we are able to obtain an AUC score of 97.6\% on data from the NGTS survey, showing that such techniques are effective for ground based photometric data with complex window functions compared to the space-based photometry typically used as a test case. We demonstrate that machine learning methods can be effective for new and active surveys where a sizeable training set does not yet exist by utilising injected synthetic transits. Such simulations further allow a degree of testing and interpretability of the model. 

Improvements to our method are possible, through inclusion of stellar data from the GAIA satellite, increasing the size of the training set, taking more account of outliers and the parameter space viewed by the model, and potentially by exploring alternate inherently probabilistic classifiers. In particular, understanding potential biases and improving reliability on individual candidates is crucial for fully taking advantage of automated methods.

We also present a publicly accessible code, \texttt{autovet}, which can calculate features from lightcurves and act as a wrapper for various \texttt{scikit-learn} machine learning implementations. Similar techniques are becoming increasingly prevalent as surveys produce increasing quantities of data, and will be crucial in maximising the scientific return of missions such as TESS and PLATO.

\section*{Acknowledgements}
This publication is based on data collected under the NGTS project at the ESO La Silla Paranal Observatory. The NGTS instrument and operations are funded by the consortium institutes and by the UK Science and Technology Facilities Council (STFC; project reference ST/M001962/1). DJA, DP, PJW and RGW are supported by an STFC consolidated grant (ST/P000495/1).
MNG is supported by the UK Science and Technology Facilities Council (STFC) award reference 1490409 as well as the Isaac Newton Studentship.
JSJ acknowledges support by FONDECYT grant 1161218 and partial support by CATA-Basal (PB06, CONICYT) We thank the anonymous referee for helpful comments which improved the manuscript.

\bibliography{papers010518}

\begin{thebibliography}{}

\bibitem[\protect\citeauthoryear{Almenara et~al.,}{Almenara
  et~al.}{2009}]{Almenara:2009fb}
Almenara J.~M.  et~al., 2009, Astronomy and Astrophysics, 506, 337

\bibitem[\protect\citeauthoryear{Alonso et~al.,}{Alonso
  et~al.}{2004}]{Alonso:2004ii}
Alonso R.  et~al., 2004, The Astrophysical Journal, 613, L153

\bibitem[\protect\citeauthoryear{Armstrong et~al.,}{Armstrong
  et~al.}{2016}]{Armstrong:2016br}
Armstrong D.~J.  et~al., 2016, Monthly Notices of the Royal Astronomical
  Society, 456, 2260

\bibitem[\protect\citeauthoryear{Armstrong, Pollacco \& Santerne}{Armstrong
  et~al.}{2017}]{Armstrong:2017cp}
Armstrong D.~J.,  Pollacco D.,    Santerne A.,  2017, Monthly Notices of the
  Royal Astronomical Society, 465, 2634

\bibitem[\protect\citeauthoryear{Bakos, Noyes, Kovacs, Stanek, Sasselov \&
  Domsa}{Bakos et~al.}{2004}]{Bakos:2004gx}
Bakos G.,  Noyes R.~W.,  Kovacs G.,  Stanek K.~Z.,  Sasselov D.~D.,    Domsa
  I.,  2004, Publications of the Astronomical Society of the Pacific, 116, 266

\bibitem[\protect\citeauthoryear{Bakos et~al.,}{Bakos
  et~al.}{2013}]{Bakos:2013fc}
Bakos G.~A.  et~al., 2013, Publications of the Astronomical Society of the
  Pacific, 125, 154

\bibitem[\protect\citeauthoryear{Bakos et~al.,}{Bakos
  et~al.}{2007}]{Bakos:2007tt}
Bakos G.~A.  et~al., 2007, The Astrophysical Journal, 656, 552

\bibitem[\protect\citeauthoryear{Bayliss et~al.,}{Bayliss
  et~al.}{2017}]{Bayliss:2017bw}
Bayliss D.  et~al., 2017, Monthly Notices of the Royal Astronomical Society

\bibitem[\protect\citeauthoryear{Blomme et~al.,}{Blomme
  et~al.}{2010}]{Blomme:2010bq}
Blomme J.  et~al., 2010, The Astrophysical Journal, 713, L204

\bibitem[\protect\citeauthoryear{Borucki et~al.,}{Borucki
  et~al.}{2010}]{Borucki:2010dn}
Borucki W.~J.  et~al., 2010, Science, 327, 977

\bibitem[\protect\citeauthoryear{Bouchy et~al.,}{Bouchy
  et~al.}{2005}]{Bouchy:2005wb}
Bouchy F.  et~al., 2005, Astronomy and Astrophysics, 444, L15

\bibitem[\protect\citeauthoryear{Brahm et~al.,}{Brahm
  et~al.}{2017}]{Brahm:2017uc}
Brahm R.  et~al., 2017, eprint arXiv:1707.07093

\bibitem[\protect\citeauthoryear{Breiman}{Breiman}{2001}]{Breiman:fb}
Breiman L.,  2001, Machine Learning, 45, 5

\bibitem[\protect\citeauthoryear{Brett, West \& Wheatley}{Brett
  et~al.}{2004}]{Brett:2004cr}
Brett D.~R.,  West R.~G.,    Wheatley P.~J.,  2004, Monthly Notices of the
  Royal Astronomical Society, 353, 369

\bibitem[\protect\citeauthoryear{Brink, Richards, Poznanski, Bloom, Rice,
  Negahban \& Wainwright}{Brink et~al.}{2013}]{Brink:2013hv}
Brink H.,  Richards J.~W.,  Poznanski D.,  Bloom J.~S.,  Rice J.,  Negahban S.,
     Wainwright M.,  2013, Monthly Notices of the Royal Astronomical Society,
  435, 1047

\bibitem[\protect\citeauthoryear{Cabrera et~al.,}{Cabrera
  et~al.}{2017}]{Cabrera:2017fo}
Cabrera J.  et~al., 2017, Astronomy and Astrophysics, 606, A75

\bibitem[\protect\citeauthoryear{Carrasco et~al.,}{Carrasco
  et~al.}{2015}]{Carrasco:2015kh}
Carrasco D.  et~al., 2015, Astronomy and Astrophysics, 584, A44

\bibitem[\protect\citeauthoryear{Carrasco~Kind \& Brunner}{Carrasco~Kind \&
  Brunner}{2014}]{CarrascoKind:2014gb}
Carrasco~Kind M.,  Brunner R.~J.,  2014, Monthly Notices of the Royal
  Astronomical Society, 438, 3409

\bibitem[\protect\citeauthoryear{Charbonneau, Brown, Latham \&
  Mayor}{Charbonneau et~al.}{2000}]{Charbonneau:2000fh}
Charbonneau D.,  Brown T.~M.,  Latham D.~W.,    Mayor M.,  2000, The
  Astrophysical Journal, 529, L45

\bibitem[\protect\citeauthoryear{Collier~Cameron et~al.,}{Collier~Cameron
  et~al.}{2007a}]{CollierCameron:2007ew}
Collier~Cameron A.  et~al., 2007a, Monthly Notices of the Royal Astronomical
  Society, 375, 951

\bibitem[\protect\citeauthoryear{Collier~Cameron et~al.,}{Collier~Cameron
  et~al.}{2006}]{2006MNRAS.373..799C}
Collier~Cameron A.  et~al., 2006, Monthly Notices of the Royal Astronomical
  Society, 373, 799

\bibitem[\protect\citeauthoryear{Collier~Cameron et~al.,}{Collier~Cameron
  et~al.}{2007b}]{CollierCameron:2007cy}
Collier~Cameron A.  et~al., 2007b, Monthly Notices of the Royal Astronomical
  Society, 380, 1230

\bibitem[\protect\citeauthoryear{Coughlin et~al.,}{Coughlin
  et~al.}{2016}]{Coughlin:2016cm}
Coughlin J.~L.  et~al., 2016, The Astrophysical Journal Supplement Series, 224,
  12

\bibitem[\protect\citeauthoryear{Crossfield et~al.,}{Crossfield
  et~al.}{2016}]{Crossfield:2016ip}
Crossfield I. J.~M.  et~al., 2016, The Astrophysical Journal Supplement Series,
  226, 7

\bibitem[\protect\citeauthoryear{Debosscher, Blomme, Aerts \&
  De~Ridder}{Debosscher et~al.}{2011}]{Debosscher:2011kz}
Debosscher J.,  Blomme J.,  Aerts C.,    De~Ridder J.,  2011, Astronomy and
  Astrophysics, 529, A89

\bibitem[\protect\citeauthoryear{Delrez et~al.,}{Delrez
  et~al.}{2014}]{Delrez:2014iv}
Delrez L.  et~al., 2014, Astronomy and Astrophysics, 563, A143

\bibitem[\protect\citeauthoryear{D{\'\i}az, Almenara, Santerne, Moutou,
  Lethuillier \& Deleuil}{D{\'\i}az et~al.}{2014}]{Diaz:2014kd}
D{\'\i}az R.~F.,  Almenara J.~M.,  Santerne A.,  Moutou C.,  Lethuillier A.,
  Deleuil M.,  2014, Monthly Notices of the Royal Astronomical Society, 441,
  983

\bibitem[\protect\citeauthoryear{Dittmann et~al.,}{Dittmann
  et~al.}{2017}]{Dittmann:2017df}
Dittmann J.~A.  et~al., 2017, Nature, 544, 333

\bibitem[\protect\citeauthoryear{Eyer \& Blake}{Eyer \&
  Blake}{2005}]{Eyer:2005ce}
Eyer L.,  Blake C.,  2005, Monthly Notices of the Royal Astronomical Society,
  358, 30

\bibitem[\protect\citeauthoryear{Farrell, Murphy \& Lo}{Farrell
  et~al.}{2015}]{Farrell:2015je}
Farrell S.~A.,  Murphy T.,    Lo K.~K.,  2015, The Astrophysical Journal
  Letters, 813, 28

\bibitem[\protect\citeauthoryear{Fawcett}{Fawcett}{2006}]{Fawcett:2006gr}
Fawcett T.,  2006, Pattern Recognition Letters, 27, 861

\bibitem[\protect\citeauthoryear{Fressin et~al.,}{Fressin
  et~al.}{2013}]{Fressin:2013df}
Fressin F.  et~al., 2013, The Astrophysical Journal, 766, 81

\bibitem[\protect\citeauthoryear{{G{\"u}nther}, {Queloz}, {Demory} \&
  {Bouchy}}{{G{\"u}nther} et~al.}{2017}]{Guenther:2017a}
{G{\"u}nther} M.~N.,  {Queloz} D.,  {Demory} B.-O.,    {Bouchy} F.,  2017,
  Monthly Notices of the Royal Astronomical Society, 465, 3379

\bibitem[\protect\citeauthoryear{G{\"u}nther et~al.,}{G{\"u}nther
  et~al.}{2017}]{Gunther:2017kd}
G{\"u}nther M.~N.  et~al., 2017, Monthly Notices of the Royal Astronomical
  Society, 472, 295

\bibitem[\protect\citeauthoryear{Hanke, Halchenko, Sederberg, Hanson, Haxby \&
  Pollmann}{Hanke et~al.}{2009}]{Hanke:2009bm}
Hanke M.,  Halchenko Y.~O.,  Sederberg P.~B.,  Hanson S.~J.,  Haxby J.~V.,
  Pollmann S.,  2009, Neuroinformatics, 7, 37

\bibitem[\protect\citeauthoryear{{Hartman}, {Bakos} \& {Torres}}{{Hartman}
  et~al.}{2011}]{Hartman:2011}
{Hartman} J.~D.,  {Bakos} G.~{\'A}.,    {Torres} G.,  2011, in European
  Physical Journal Web of Conferences. p.~2002

\bibitem[\protect\citeauthoryear{Hellier et~al.,}{Hellier
  et~al.}{2017}]{Hellier:2017jv}
Hellier C.  et~al., 2017, Monthly Notices of the Royal Astronomical Society,
  465, 3693

\bibitem[\protect\citeauthoryear{Hellier et~al.,}{Hellier
  et~al.}{2014}]{Hellier:2014dqa}
Hellier C.  et~al., 2014, Monthly Notices of the Royal Astronomical Society,
  440, 1982

\bibitem[\protect\citeauthoryear{Huang, Ma, Zhao \& Lu}{Huang
  et~al.}{2017}]{Huang:2017eb}
Huang C.,  Ma Y.-h.,  Zhao H.-b.,    Lu X.-p.,  2017, Chinese Astronomy and
  Astrophysics, 41, 549

\bibitem[\protect\citeauthoryear{Kohonen}{Kohonen}{1982}]{Kohonen:1982dy}
Kohonen T.,  1982, Biological Cybernetics, 43, 59

\bibitem[\protect\citeauthoryear{Kovacs, Zucker \& Mazeh}{Kovacs
  et~al.}{2002}]{Kovacs:2002ho}
Kovacs G.,  Zucker S.,    Mazeh T.,  2002, Astronomy and Astrophysics, 391, 369

\bibitem[\protect\citeauthoryear{Kreidberg}{Kreidberg}{2015}]{Kreidberg:2015di}
Kreidberg L.,  2015, Publications of the Astronomical Society of the Pacific,
  127, 1161

\bibitem[\protect\citeauthoryear{Latham et~al.,}{Latham
  et~al.}{2009}]{Latham:2009bx}
Latham D.~W.  et~al., 2009, The Astrophysical Journal, 704, 1107

\bibitem[\protect\citeauthoryear{LeCun, Bengio \& Hinton}{LeCun
  et~al.}{2015}]{LeCun:2015dt}
LeCun Y.,  Bengio Y.,    Hinton G.,  2015, Nature, 521, 436

\bibitem[\protect\citeauthoryear{Liu, Deng, Wang \& Wang}{Liu
  et~al.}{2017}]{Liu:2017dv}
Liu C.,  Deng N.,  Wang J. T.~L.,    Wang H.,  2017, The Astrophysical Journal
  Letters, 843, 104

\bibitem[\protect\citeauthoryear{Liu, Ting \& Zhou}{Liu et~al.}{2008}]{Liu:kh}
Liu F.~T.,  Ting K.~M.,    Zhou Z.-H.,  2008, in 2008 Eighth IEEE International
  Conference on Data Mining (ICDM). IEEE, pp 413--422

\bibitem[\protect\citeauthoryear{McCauliff et~al.,}{McCauliff
  et~al.}{2015}]{McCauliff:2015fb}
McCauliff S.~D.  et~al., 2015, The Astrophysical Journal, 806, 6

\bibitem[\protect\citeauthoryear{McCormac, Pollacco, Skillen, Faedi, Todd \&
  Watson}{McCormac et~al.}{2013}]{McCormac:2013hc}
McCormac J.,  Pollacco D.,  Skillen I.,  Faedi F.,  Todd I.,    Watson C.~A.,
  2013, Publications of the Astronomical Society of the Pacific, 125, 548

\bibitem[\protect\citeauthoryear{McCormac et~al.,}{McCormac
  et~al.}{2017}]{McCormac:2017ju}
McCormac J.  et~al., 2017, Publications of the Astronomical Society of the
  Pacific, 129, 025002

\bibitem[\protect\citeauthoryear{McCullough, Stys, Valenti, Fleming, Janes \&
  Heasley}{McCullough et~al.}{2005}]{McCullough:2005do}
McCullough P.~R.,  Stys J.~E.,  Valenti J.~A.,  Fleming S.~W.,  Janes K.~A.,
  Heasley J.~N.,  2005, Publications of the Astronomical Society of the
  Pacific, 117, 783

\bibitem[\protect\citeauthoryear{Mahabal et~al.,}{Mahabal
  et~al.}{2008}]{Mahabal:2008if}
Mahabal A.  et~al., 2008, Astronomische Nachrichten, 329, 288

\bibitem[\protect\citeauthoryear{Mandel \& Agol}{Mandel \&
  Agol}{2002}]{Mandel:2002bb}
Mandel K.,  Agol E.,  2002, The Astrophysical Journal, 580, L171

\bibitem[\protect\citeauthoryear{Masci, Hoffman, Grillmair \& Cutri}{Masci
  et~al.}{2014}]{Masci:2014bk}
Masci F.~J.,  Hoffman D.~I.,  Grillmair C.~J.,    Cutri R.~M.,  2014, The
  Astronomical Journal, 148, 21

\bibitem[\protect\citeauthoryear{Mislis, Bachelet, Alsubai, Bramich \&
  Parley}{Mislis et~al.}{2015}]{Mislis:2015cj}
Mislis D.,  Bachelet E.,  Alsubai K.~A.,  Bramich D.~M.,    Parley N.,  2015,
  Monthly Notices of the Royal Astronomical Society, 455, 626

\bibitem[\protect\citeauthoryear{Morton}{Morton}{2012}]{Morton:2012bv}
Morton T.~D.,  2012, The Astrophysical Journal, 761, 6

\bibitem[\protect\citeauthoryear{Morton, Bryson, Coughlin, Rowe, Ravichandran,
  Petigura, Haas \& Batalha}{Morton et~al.}{2016}]{Morton:2016ka}
Morton T.~D.,  Bryson S.~T.,  Coughlin J.~L.,  Rowe J.~F.,  Ravichandran G.,
  Petigura E.~A.,  Haas M.~R.,    Batalha N.~M.,  2016, The Astrophysical
  Journal, 822, 86

\bibitem[\protect\citeauthoryear{Nun, Pichara, Protopapas \& Kim}{Nun
  et~al.}{2014}]{Nun:2014kv}
Nun I.,  Pichara K.,  Protopapas P.,    Kim D.-W.,  2014, The Astrophysical
  Journal, 793, 23

\bibitem[\protect\citeauthoryear{Pearson, Palafox \& Griffith}{Pearson
  et~al.}{2018}]{Pearson:2018kf}
Pearson K.~A.,  Palafox L.,    Griffith C.~A.,  2018, Monthly Notices of the
  Royal Astronomical Society, 474, 478

\bibitem[\protect\citeauthoryear{Pedregosa et~al.,}{Pedregosa
  et~al.}{2011}]{Pedregosa:2011tv}
Pedregosa F.  et~al., 2011, Journal of Machine Learning Research, 12, 2825

\bibitem[\protect\citeauthoryear{Pepper et~al.,}{Pepper
  et~al.}{2007}]{Pepper:2007ja}
Pepper J.  et~al., 2007, Publications of the Astronomical Society of the
  Pacific, 119, 923

\bibitem[\protect\citeauthoryear{Pollacco et~al.,}{Pollacco
  et~al.}{2006}]{Pollacco:2006gb}
Pollacco D.~L.  et~al., 2006, Publications of the Astronomical Society of the
  Pacific, 118, 1407

\bibitem[\protect\citeauthoryear{Richards et~al.,}{Richards
  et~al.}{2011}]{Richards:2011ji}
Richards J.~W.  et~al., 2011, The Astrophysical Journal, 733, 10

\bibitem[\protect\citeauthoryear{Richards, Starr, Miller, Bloom, Butler, Brink
  \& Crellin-Quick}{Richards et~al.}{2012}]{Richards:2012ea}
Richards J.~W.,  Starr D.~L.,  Miller A.~A.,  Bloom J.~S.,  Butler N.~R.,
  Brink H.,    Crellin-Quick A.,  2012, The Astrophysical Journal Supplement
  Series, 203, 32

\bibitem[\protect\citeauthoryear{Ricker et~al.,}{Ricker
  et~al.}{2014}]{Ricker:2014fy}
Ricker G.~R.  et~al., 2014, in SPIE Astronomical Telescopes + Instrumentation.
  SPIE, p. 914320

\bibitem[\protect\citeauthoryear{Santerne et~al.,}{Santerne
  et~al.}{2015}]{Santerne:2015bb}
Santerne A.  et~al., 2015, Monthly Notices of the Royal Astronomical Society,
  451, 2337

\bibitem[\protect\citeauthoryear{Shallue \& Vanderburg}{Shallue \&
  Vanderburg}{2017}]{Shallue:2017vy}
Shallue C.~J.,  Vanderburg A.,  2017, eprint arXiv:1712.05044

\bibitem[\protect\citeauthoryear{Shporer et~al.,}{Shporer
  et~al.}{2017}]{Shporer:2017hv}
Shporer A.  et~al., 2017, The Astrophysical Journal, 847, L18

\bibitem[\protect\citeauthoryear{Smith et~al.,}{Smith
  et~al.}{2012}]{Smith:2012ji}
Smith J.~C.  et~al., 2012, Publications of the Astronomical Society of the
  Pacific, 124, 1000

\bibitem[\protect\citeauthoryear{Stumpe et~al.,}{Stumpe
  et~al.}{2012}]{Stumpe:2012bj}
Stumpe M.~C.  et~al., 2012, Publications of the Astronomical Society of the
  Pacific, 124, 985

\bibitem[\protect\citeauthoryear{Sullivan et~al.,}{Sullivan
  et~al.}{2015}]{Sullivan:2015ey}
Sullivan P.~W.  et~al., 2015, The Astrophysical Journal Letters, p.~77

\bibitem[\protect\citeauthoryear{Tamuz, Mazeh \& Zucker}{Tamuz
  et~al.}{2005}]{Tamuz:2005hi}
Tamuz O.,  Mazeh T.,    Zucker S.,  2005, Monthly Notices of the Royal
  Astronomical Society, 356, 1466

\bibitem[\protect\citeauthoryear{Thompson et~al.,}{Thompson
  et~al.}{2017}]{Thompson:2017wu}
Thompson S.~E.  et~al., 2017, eprint arXiv:1710.06758

\bibitem[\protect\citeauthoryear{Thompson, Mullally, Coughlin, Christiansen,
  Henze, Haas \& Burke}{Thompson et~al.}{2015}]{Thompson:2015ic}
Thompson S.~E.,  Mullally F.,  Coughlin J.,  Christiansen J.~L.,  Henze C.~E.,
  Haas M.~R.,    Burke C.~J.,  2015, The Astrophysical Journal Letters, p.~46

\bibitem[\protect\citeauthoryear{Torres et~al.,}{Torres
  et~al.}{2015}]{2015ApJ...800...99T}
Torres G.  et~al., 2015, The Astrophysical Journal, 800, 99

\bibitem[\protect\citeauthoryear{Wheatley et~al.,}{Wheatley
  et~al.}{2017}]{Wheatley:2017dm}
Wheatley P.~J.  et~al., 2017, Monthly Notices of the Royal Astronomical Society

\end{thebibliography}
\bibliographystyle{mn2e_fix}

\end{document}